\documentclass[useAMS,usenatbib]{mn2e}
\usepackage{color}
\usepackage{times}
\usepackage{graphicx}
\usepackage{amsmath}
\usepackage{amssymb}
\newcommand{\aap}{A\&A}

\newcommand{\aj}{AJ}
\newcommand{\apj}{ApJ}
\newcommand{\apjs}{ApJS}
\newcommand{\aapr}{A\&ARV}

\title[Extinction correction for NIR galaxy parameters]{The Effect of Dust Extinction
on the Observed Properties of Galaxies in the Near-Infrared.}
\date{}
\author[Ihab F. Riad et al.]{Ihab F. Riad,$^1$ Ren\'ee C. Kraan-Korteweg$^1$ and Patrick
A. Woudt$^1$ \\$^1$Department of Astronomy, University of Cape Town, Private Bag, X3
Rondebosch 7701, South Africa}
\begin{document}
\maketitle
\begin{abstract}
Galaxies behind the Milky Way suffer size reduction and dimming due to their
obscuration
by dust in the disk of our Galaxy. The degree of obscuration is wavelength dependent. It
decreases towards longer wavelengths. Compared to the optical, the Near InfraRed (NIR)
$K_s$~band extinction is only $\approx10\%$ that of the $B$ band. This makes NIR surveys
well suited for galaxy surveys close to the Galactic Plane where extinction is severe.

While Galactic obscuration is less prominent in the NIR it is not negligible. In this
paper we derive empirical relations to correct isophotal radii and magnitudes of galaxies
observed in the NIR for foreground absorption. We simulate extinction in the $J$, $H$ and
$K_s$ bands on 64 (unobscured) galaxies from the 2MASS Large Galaxy Atlas \citep{jarrett}.
We propose two methods for the extinction correction, the first is optimized to provide
the most accurate correction and the second provides a convenient statistical correction
that works adequately in lower extinction regions. The optimized correction utilizes the
galaxy surface brightness, either the disk central surface brightness, $\mu_0$, or the
combined disk plus bulge central surface brightness, elliptical and disk/spiral Hubble
types. A detailed comparison between the different methods and their accuracy is provided.
\end{abstract}
\section{Introduction}
The effect of dust extinction on point-like objects, stars, is linearly related to
extinction. If, for example, the observed magnitude of a star is $K_s=6\fm0$ and it
suffers an extinction of $A_{K_s}=1\fm5$, its intrinsic magnitude will be
$K_s^\circ=4\fm5$. Extended objects, like galaxies, suffer a further dimming due to the
loss of their fainter outer regions in the sky background. An obscured galaxy appears
smaller and fainter in the sky than it really is. Corrections for this extra dimming are
non-linear as was shown by \citet{{ft1981},{haus1987}} and \citet{cameron} in the optical.
\citet{taka} derived corrections to the $K_s$ band isophotal magnitudes. Their extinction
correction study was initiated to correct the isophotal magnitudes for galaxies observed
in the vicinity of the giant radio galaxy PKS 1343-601 centred at
$(\ell,b)=(309\fdg7,1\fdg7)$. The average extinction in their region was
$A_{K_s}\lesssim1\fm1$. From their study they showed that the isophotal magnitude
correction can be approximated by a linear relation for extinction levels
$A_{K_s}\lesssim1\fm0$. Their work also showed that the different morphological types of
galaxies are affected differently by extinction. Galaxies that have an exponential light
profile bulge (late type) need larger corrections than those with a de~Vaucouleurs profile
(early type), or those with a bulge+disk light profile.

Statistical results and conclusions derived from magnitude-limited and radius-limited
galaxy catalogues are unreliable if they are not corrected for these extinction effects.
Applying extinction corrections to galaxy catalogues like the 2MASS Extended Source
Catalog (2MASX) \citep{xsc} will imply fainter completeness magnitudes than the currently
quoted values. Applying extinction corrections to optical catalogues for galaxies observed
in the Zone of Avoidance (ZOA) helped reduce the optical ZOA from the region with
extinction $A_B\lesssim1\fm0$, to the region with $A_B\lesssim3\fm0$
\citep{kraan-lahav2000}. This reduction was a result of including those galaxies that did
not make it into the diameter limited catalogues if they were not corrected for
extinction.

Extinction magnitude corrections are crucial when relating spiral galaxy magnitudes to
their velocity width for the Tully-Fisher (TF) relation. The here derived NIR extinction
corrections will prove invaluable for e.g. the ongoing whole-sky 2MASS Tully Fisher Survey
\citep{masters}, especially for galaxies found in the ZOA. Application of NIR magnitude
extinction corrections will also improve photometric redshift estimations \citep{prshift}
for galaxies that are observed in the ZOA.

Motivated by an ongoing NIR imaging survey to map the crossing of the Great Attractor wall
across the ZOA where extinction levels are severe, we initiated a study of the effect of
Galactic extinction on galaxies imaged in the $J(1.2\mu m)$, $H(1.6\mu m)$ and $K_s(2.2\mu
m)$ bands. This will allow us to correct their observed isophotal-radii and magnitudes for
extinction. In our current ZOA survey along the Norma wall (Riad et al., in
preparation) we noticed, that $\approx 85\%$ of the galaxies in our sample were found in
regions with an extinction $A_{J}\lesssim1\fm0$, while $\approx 14\%$ were found in
regions with an extinction $1\fm0\lesssim A_{J}\lesssim3\fm0$. Only $1\%$ of the galaxies
were found in regions with an extinction $A_J>3\fm0$. For the $H$ and $K_s$ band $\approx
94\%$ and $98\%$ respectively of the galaxies were found in regions with
$A_{H,K_s}\lesssim1\fm0$, while $\approx 5\%$ and $1\%$ of the galaxies were obscured by
$1\fm0\lesssim A_{H,K_s}\lesssim3\fm0$. Only a handfull of the galaxies were found in
regions with $A_{H,K_s}>3\fm0$. We therefore limited the simulation to the extinction
range $A_{J,H,K_s}=0\fm0-3\fm0$. In this work and for the purpose of estimating the
extinction suffered by galaxies in our ZOA survey we used the general extinction law,
\begin{eqnarray}
\frac{A_V}{E(B-V)}&=&R_V \nonumber.
\end{eqnarray}
Where $E(B-V)$ is the colour reddening derived from the \citet{schlegel} reddening maps
and $A_V$ is the extinction in the optical $V$ band. A typical value for $R_V$ is $3.1$
\citep{cardelli}. Extinction in the NIR $J$, $H$ and $K_s$ passbands was derived using the
parametrization given by \citet{cardelli}, see Eqs. \ref{extj} - \ref{extk}.
\begin{eqnarray}
A_J&=&0.874\,E(B-V),\label{extj}\\
A_H&=&0.549\,E(B-V),\\
A_{K_s}&=&0.365\,E(B-V).\label{extk}
\end{eqnarray}

In this paper, we derive NIR corrections to isophotal magnitudes and radii for galaxies
obscured by the Milky Way. Section 2 describes the data set and method. We describe two
methods to apply the corrections (Sects. 2.2 and 2.3). In Sect. 3 we provide a brief
comparison between the different methods, and discuss their respective reliability.
\section{Data and method}
We selected 64 galaxies from the 2MASS Large Galaxy Atlas (LGA) \citep{jarrett} to
simulate the effect of extinction on isophotal-radii and magnitude for galaxies observed
in the NIR. 2MASS is an All-Sky NIR survey in the bands $J(1.2\:\mu m)$, $H(1.6\:\mu m)$
and $K_s(2.2\:\mu m)$. Galaxies were selected in such a way that they are minimally
affected by contamination from neighbouring sources, and give a fair representation of all
morphological types. The sample includes 25 elliptical and lenticular ({\bf
E/S0}), and 39 spiral ({\bf S}) galaxies. Half of the galaxies in the spiral galaxy sample
are barred with some of them having ring features. The selected galaxies cover a wide
range in galaxy size and brightness. The apparent radii range between
$r_{20}=34\arcsec-1614\arcsec$ where $r_{20}$ is the isophotal radius at the surface
brightness level $\mu_K=20$ mag/arcsec$^2$. The apparent isophotal magnitude range covered
by our sample is $K_{20}=1\fm04 - 10\fm77$, where $K_{20}$ is the integrated magnitude
within $r_{20}$. The galaxies in our sample suffered minimal obscuration ranging between
$A_{K_s}=0\fm01-0\fm07$, see last column in Table \ref{tgalaxies}. The list of selected
galaxies is given in Table \ref{tgalaxies}. This table lists the most common name of the
galaxy, followed by the morphological type, $r_{20}$, $K_{20}$, inclination $(b/a)$ and
position angle (PA) in the $K_s$ band. All the data are taken from \citet{jarrett}. In the
table we also give the disk central surface brightness $\mu_0$ \citep{freeman} and the
central surface brightness $\mu_c$ in the $K_s$ band. The last column lists the mean
extinction $A_{K_s}$. Extinction in the $K_s$ band was derived from the \citet{schlegel}
reddening maps and Eqn. \ref{extk}. The surface brightness profiles for the galaxies are
taken from the LGA\footnote[1]{The surface brightness profile for galaxies in the LGA are
found in http://irsa.ipac.caltech.edu/applications/2MASS/LGA/} (given in tabular format).

\begin{table*}
\begin{minipage}{130mm}
\begin{center}
\caption{The sample of galaxies used for the study of extinction effects on the
NIR.}\label{tgalaxies}
\begin{scriptsize}
\begin{tabular}{lllrcccccc}
\hline
No.& Galaxy & Morphology & $r_{20}$ & $K_{20}$ & $b/a$ &
PA&$\mu_0$&$\mu_c$&$A_{K_s}$\\
&&&[$\arcsec$]&[mag]&&[$^\circ$]&[mag/arcsec$^2$]&[mag/arcsec$^2$]&[mag]\\
\hline
  1&    NGC4697   &E6     &123.8&6.502&0.63&$+$67.5&16.10&12.88&0.011\\
  2&    M86       &S03/E3 &151.4&6.283&0.67&$-$55.0&16.37&13.21&0.011\\
  3&    M60       &E2     &146.6&5.825&0.81&$-$72.5&15.83&13.06&0.010\\
  4&    M59       &E5     &109.0&6.866&0.65&$-$15.0&16.11&12.67&0.012\\
  5&    M84       &E1     &115.0&6.347&0.92&$-$57.5&16.12&12.93&0.015\\
  6&    M105      &E1     &108.0&6.362&0.85&$+$67.5&15.92&12.65&0.009\\
  7&    NGC584    &E4     &88.8&7.445&0.62&$+$62.5&16.18&13.05&0.015\\
  8&    NGC720    &E5     &90.5&7.396&0.55&$-$40.0&16.14&13.63&0.006\\
  9&    NGC1395   &E2     &93.7&7.024&0.82&$-$87.5&16.10&13.29&0.008\\
  10&   NGC1407   &E0     &100.2&6.855&0.95&$+$60.0&16.25&13.62&0.025\\
  11&   NGC4365   &E3     &113.0&6.800&0.74&$+$45.0&16.32&13.36&0.008\\
  12&   NGC4473   &E5     &93.7&7.269&0.54&$-$85.0&15.96&12.89&0.010\\
  13&   NGC4494   &E1-2   &91.1&7.145&0.87&$-$07.5&16.15&13.04&0.009\\
  14&   NGC4589   &E2     &69.5&7.915&0.75&$+$92.5&16.27&13.67&0.010\\
  15&   NGC1377   &S0     &35.0&9.892&0.56&$-$86.5&16.15&15.43&0.010\\
  16&   NGC2310   &S0     &91.5&8.565&0.24&$+$47.0&15.97&14.99&0.039\\
  17&   NGC3630   &S0     &42.2&8.911&0.45&$+$37.0&15.70&13.46&0.016\\
  18&   NGC3966   &S0     &61.5&9.077&0.24&$-$72.5&15.75&14.63&0.011\\
  19&   NGC3115   &S0     &164.6&5.937&0.39&$+$45.0&15.74&12.27&0.017\\
  20&   NGC4636   &E/S0;1 &134.8&6.628&0.84&$-$37.5&16.36&13.82&0.010\\
  21&   NGC1340	  &E5     &86.0&7.546&0.62&$-$17.5&16.01&13.58&0.007\\
  22&   M49       &E2/S0(2)&179.2&5.506&0.81&$-$17.5&16.11&12.98&0.008\\
  23&   M32       &cE2    &147.4&5.139&0.87&$+$15.0&14.96&11.11&0.023\\
  24&   M87       &E+0-1;pec;Sy&136.0&5.904&0.86&$+$27.5&16.05&13.64&0.008\\
  25&   NGC855    &E	  &41.6&10.161&0.50&$+$67.0&16.71&15.97&0.026\\
  26&   NGC4244$^\star$   &SA(s)cd&157.0&8.110&0.30&$+$45.5&17.49&16.82&0.008\\
  27&   NGC55$^\star$     &SB(s)m &273.5&6.562&0.30&$-$67.0&17.12&17.03&0.005\\
  28&   NGC1073$^\star$   &SB(rs)c&57.7&9.690&0.72&$+$42.5&18.70&16.99&0.014\\
  29&   NGC247$^\star$    &SAB(s)d&141.7&8.180&0.55&$+$01.5&18.21&17.01&0.007\\
  30&   NGC24$^\star$     &SA(s)c &83.0&9.215&0.28&$+$43.5&16.97&17.03&0.007\\
  31&	M33$^\star$       &SA(s)cd&499.9&5.38&0.80&$+$50.5&17.71&14.10&0.015\\
  32&   NGC4569    &SAB(rs)ab;Sy &165.1&7.686&0.40&$-$15.0&16.35&12.83&0.017\\
  33&   NGC4216   &SAB(s)b&217.6&6.587&0.80&$+$109.5&15.76&12.97&0.012\\
  34&   M100      &SAB(s)bc;LINER&150.8&6.810&0.73&$-$72.5&17.28&13.97&0.010\\
  35&   M106      &SAB(s)b;LINER&263.9&5.598&0.49&$-$20.0&15.98&13.39&0.006\\
  36&   M51a      &SA(s)bc&197.5&5.601&0.68&$+$57.5&16.38&13.57&0.013\\
  37&   M109      &SAB(rs)bc;LINER&142.6&7.139&0.56&$+$38.0&17.17&14.02&0.010\\
  38&   NGC908    &SA(s)c&128.9&7.365&0.55&$+$87.5&16.41&14.54&0.009\\
  39&   NGC488 	  &SA(r)b&107.8&7.133&0.81&$+$05.0&16.45&13.81&0.011\\
  40&   NGC4826   &SA(rs)ab;Sy2&214.8&5.396&0.57&$-$70.0&15.24&12.38&0.015\\
  41&   NGC4594   &SA(s)a;Sy1.9&201.7&5.009&0.54&$+$87.5&16.35&11.87&0.019\\
  42&   NGC474    &SA(s)0&46.1&8.815&0.99&$+$45.0&16.42&14.05&0.013\\
  43&   NGC1532   &SB(s)b;pec;sp&143.6&6.861&0.30&$+$35.0&15.92&13.84&0.006\\
  44&   NGC7090   &SBc?;sp&136.4&8.404&0.26&$-$48.0&17.13&13.65&0.008\\
  45&   NGC2442   &SAB(s)bc;pec&126.4&7.071&0.87&$+$27.5&17.04&13.61&0.074\\
  46&   M51b      &SB0;pec&124.2&6.402&0.99&$+$27.5&15.77&12.33&0.013\\
  47&   NGC1808   &SAB(s)bc;Sy2&131.1&6.732&0.42&$-$37.0&15.78&12.77&0.011\\
  48&   NGC5005   &SAB(rs)bc;Sy2&130.8&6.501&0.42&$+$67.5&15.25&12.84&0.005\\
  49&   M88       &SA(rs)b&155.4&6.334&0.44&$-$40.0&15.56&13.24&0.014\\
  50&   NGC4527   &SAB(s)bc&141.0&7.021&0.34&$+$69.5&15.60&12.89&0.008\\
  51&   M63       &SA(rs)bc&204.2&5.728&0.58&$-$82.5&15.76&12.71&0.006\\
  52&   M98       &SA(s)ab,HII&172.8&7.012&0.34&$+$69.5&16.54&13.38&0.013\\
  53&   NGC47     &SB(rs)bc&33.8&9.988&0.55&$+$85.5&17.12&15.49&0.004\\
  54&   NGC210    &SA(s)b&54.1&8.515&0.52&$-$07.5&15.85&14.21&0.008\\
  55&   NGC628    &SA(s)c&125.3&7.187&0.86&$+$87.5&17.53&15.47&0.026\\
  56&   NGC772    &SA(s)b&105.2&7.440&0.80&$-$45.0&16.64&14.08&0.027\\
  57&   NGC1187   &SB(r)c&86.1&8.315&0.55&$+$52.5&17.18&14.28&0.008\\
  58&   NGC522    &Sc&68.9&9.389&0.14&$+$33.0&15.45&15.67&0.032\\
  59&   NGC891    &SA(s)b&225.4&5.994&0.24&$+$25.0&15.11&14.55&0.024\\
  60&   M31       &SA(s)b&1614.0&1.038&0.56&$+$45.0&14.81&11.50&0.023\\
  61&   NGC5746   &SAB(rs)b?;sp&162.2&6.927&0.26&$-$10.0&15.10&14.08&0.015\\
  62&   M94       &SA(r)ab&172.3&5.169&0.79&$+$85.0&15.50&11.70&0.007\\
  63&   NGC1300   &SB(s)bc&131.8&7.896&0.48&$-$79.0&17.63&14.26&0.011\\
  64&   M65	  &SAB(rs)a&211.6&6.113&0.32&$-$08.5&14.95&13.312&0.009\\
\hline
\end{tabular}
\end{scriptsize}
\end{center}
The table lists the morphological type, $r_{20}$, $K_{20}$, $(b/a)$ and PA, values are
taken from the LGA \citep{jarrett}. The table also lists the $K_s$ band $\mu_0$ and
$\mu_c$. Extinction in the $K_s$ band is also listed. Galaxies marked with $\star$ are
those showing the deviated correction trend explained in the next section.
\end{minipage}
\end{table*}

The method we implement here for simulating the effect of extinction on galaxies closely
follows the precepts of earlier work done by \cite{cameron} in the $B$-band.

It is difficult to define a reliable total radius and magnitude for a galaxy
\citep{jarrett}, but more straightforward to work with isophotal radii and magnitudes out
to a given radius. For this reason we only consider here corrections to isophotal radii
and the flux within that respective radius. In this paper, we define the limiting
isophotal radius $R$ as the radii to an isophote of 21.4 mag/arcsec$^2$ in the $J$, 20.6
mag/arcsec$^2$ in the $H$, and 20 mag/arcsec$^2$ in the $K_s$ band. Those values are
approximately the $1\sigma$ NIR sky background \citep{jarrett}. We also define the
integrated isophotal magnitude $m_{iso}$ as the integrated magnitude out to this radius
$R$.
\subsection{Simulating extinction}
\begin{figure}
\begin{center}
\includegraphics[width=0.23\textwidth,height=0.23\textwidth]{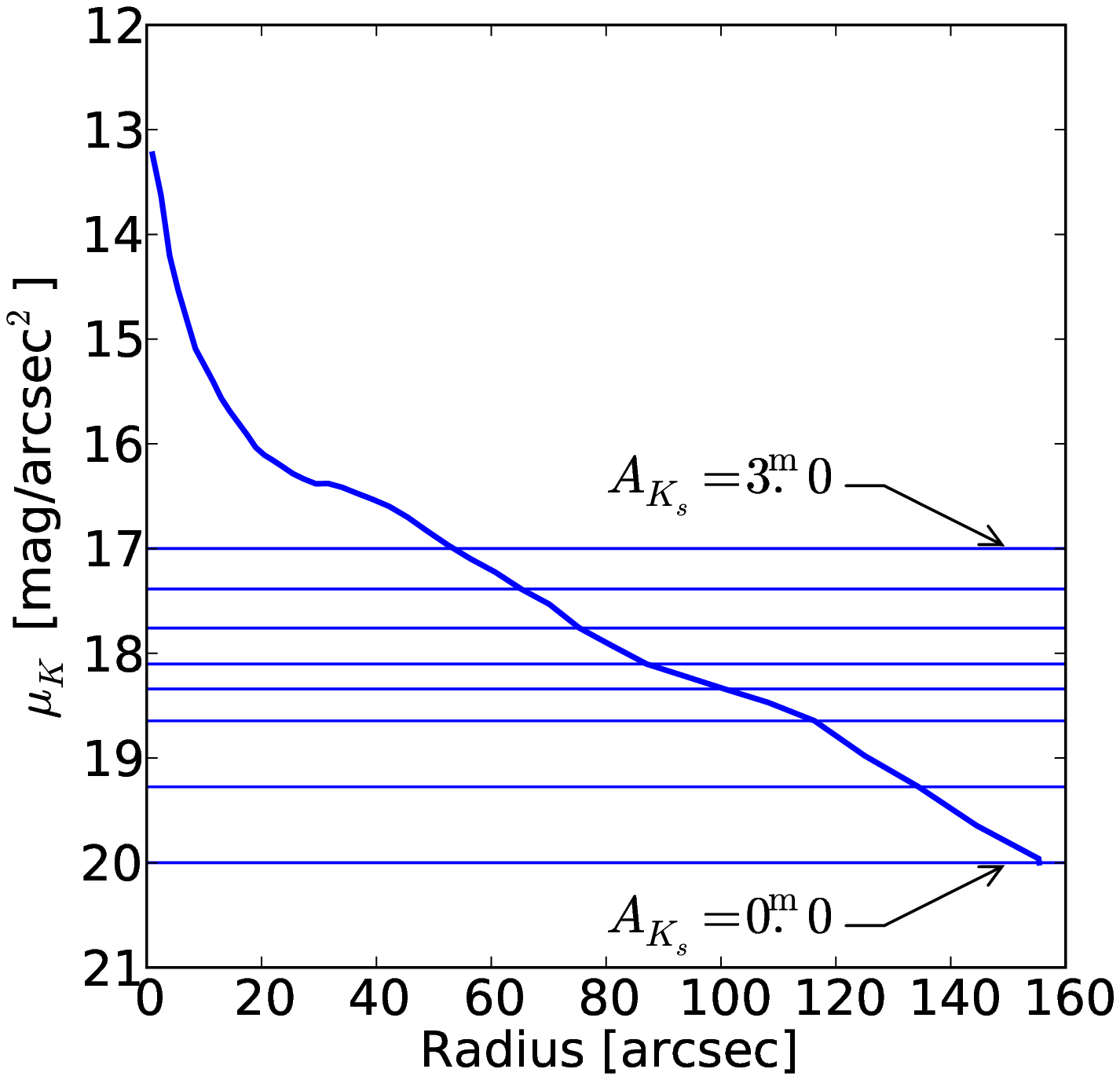}
\includegraphics[width=0.23\textwidth,height=0.23\textwidth]{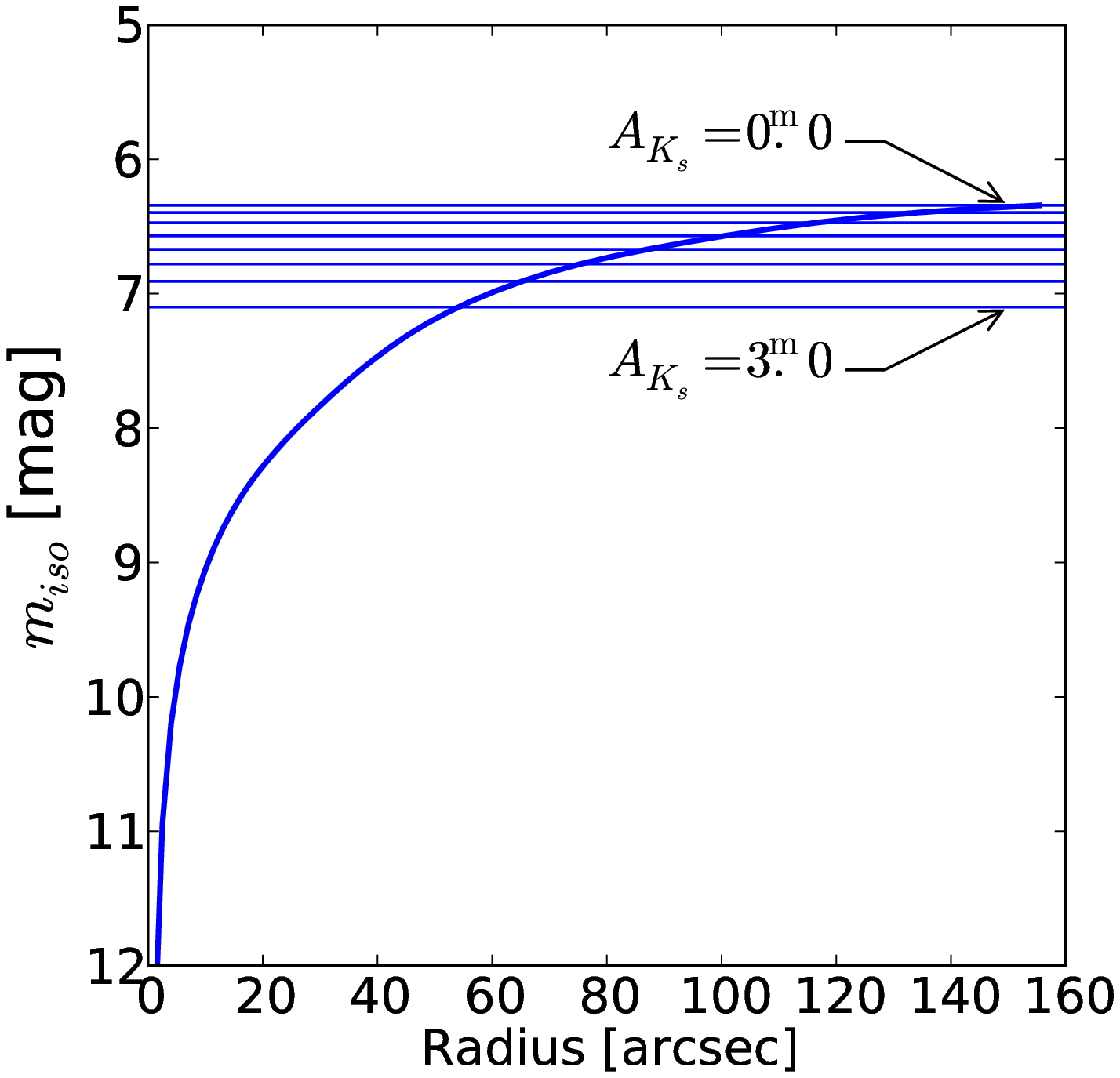}
\caption{The effect of extinction on the spiral galaxy M88. Left panel: surface brightness
profile for M88 in the $K_s$ band, with various levels of the simulated extinction in the
range $A_{K_s}=0\fm0\textrm{ to }3\fm0$. Right panel: integrated isophotal magnitude
$m_{iso}$ with extinction with the same levels of the simulated extinction.}\label{m88}
\end{center}
\end{figure}

The simulation of the effect of Galactic extinction on the apparent properties of a galaxy
was achieved by the inward displacement of the limiting isophote along the surface
brightness profile. The limiting isophote was shifted in non-uniform steps along the
galaxy light profile with each step being equivalent to a certain extinction level in the
investigated band. As an example, the left panel of Fig.~\ref{m88} shows the effect of
extinction on the surface brightness profile of the spiral galaxy M88 in the $K_s$ band.
The panel indicates increasing levels of extinction at
$A_{K_s}=0\fm0,\;0\fm7,\;1\fm4,\;1\fm7,\;1\fm9,\;2\fm2,\;2\fm6$ and $A_{K_s}=3\fm0$
applied to the galaxy. The extinction levels are represented as horizontal lines on the
plots. For each of the horizontal lines, the part of the profile lying below the line
represents the obscured part of the galaxy while the part above is what remains visible at
a given extinction. The intercept of the line with the ordinate gives the apparent
limiting isophote, while the intercept of the light profile on the abscissa shows the size
of the obscured galaxy that remains visible. The projection of the point of intercept of
the line representing $A_{K_s}=0\fm0$ with the light profile on the abscissa gives the
intrinsic radius $R^\circ$. The additional dimming caused by the loss of the faint outer
region is demonstrated by the integrated isophotal magnitude profile $m_{iso}$ shown in
Fig. \ref{m88} (right panel).

The reduced radii and magnitudes were calculated for each step and compared to the
original values. The quantities calculated as a function of extinction were $f(R)$ and
$\Delta m_{iso}$, defined as,
\begin{eqnarray}
f(R) &=&\frac{R^\circ}{R}, \nonumber\\
\Delta m_{iso}&=&m_{iso}-m_{iso}^\circ, \nonumber
\end{eqnarray}
where $m_{iso}^\circ$, $m_{iso}$, $R^\circ$ and $R$ are the intrinsic and absorbed
magnitudes and radii respectively. The different values of $f(R)$ and $\Delta m_{iso}$
corresponding to the simulated extinction values were then calculated, and fitted to the
empirical relations following the formalism given by \citet{cameron}:
\begin{eqnarray}
f(R)&=&10^{a\cdot(A_\lambda)^b}, \label{cor1-parm}\label{fit1} \\
\Delta m_{iso} &=&F\cdot(A_\lambda)^v, \label{cor2-parm} \label{fit2}
\end{eqnarray}
where $F$, $v$, $a$, $b$ are the fitting parameters, and $A_\lambda$ the extinction. This
parameter fitting was done for all galaxies in our sample. It was performed separately for
all three $J$, $H$, $K_s$ bands.

\begin{figure}
\begin{center}
\includegraphics[width=0.23\textwidth,height=0.23\textwidth]{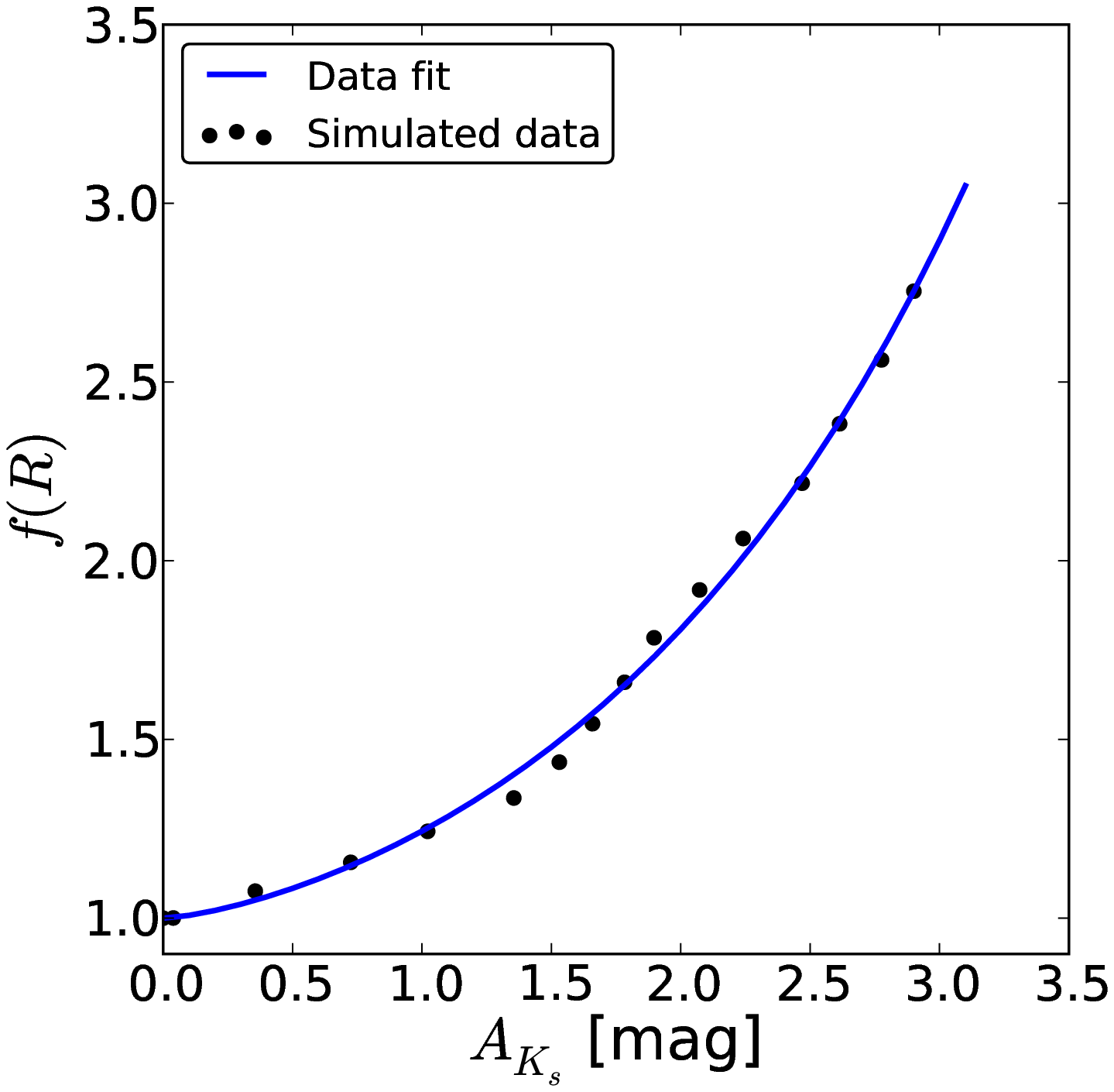}
\includegraphics[width=0.23\textwidth,height=0.23\textwidth]{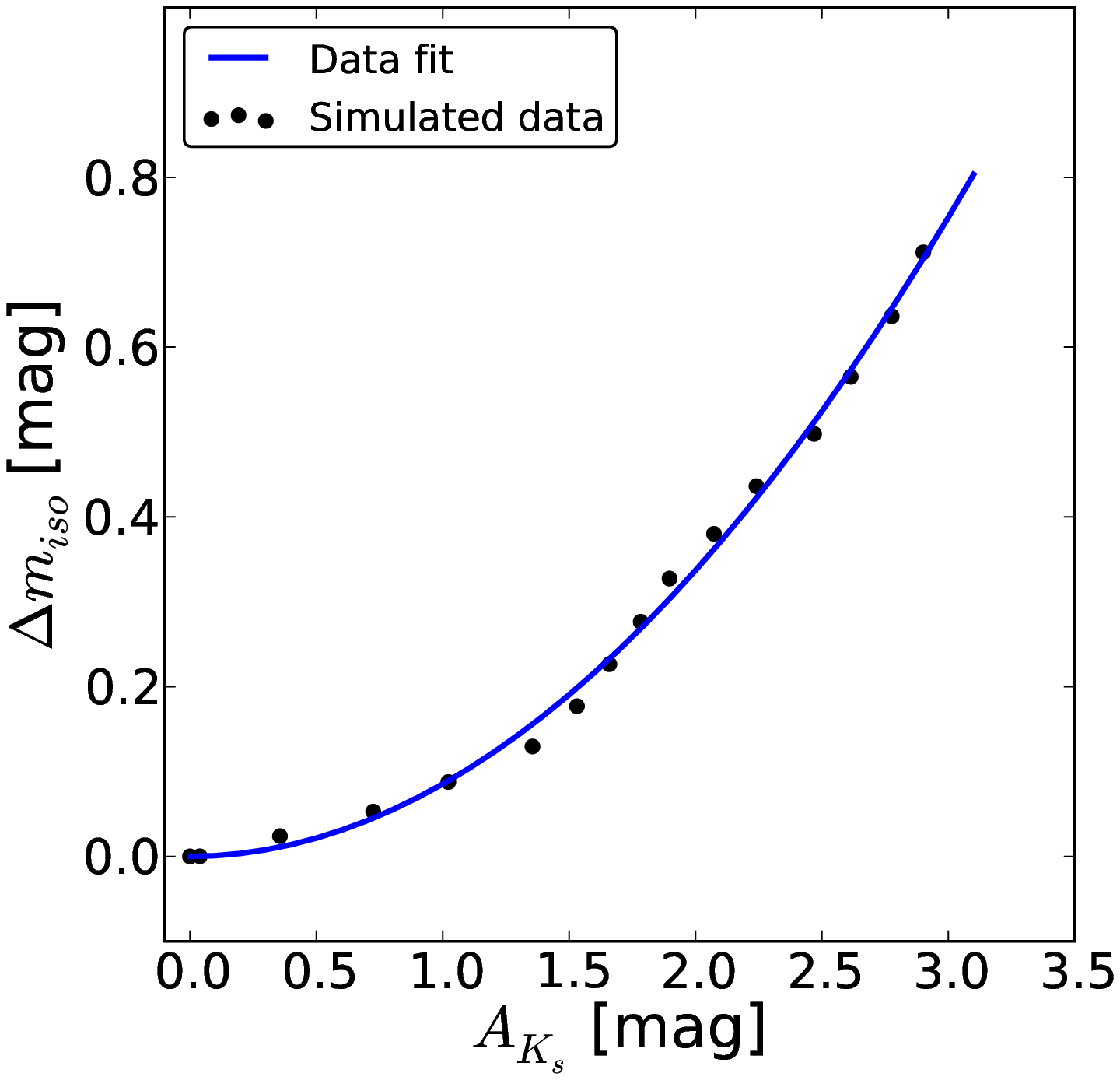}
\caption{The calculated $f(R)$ (left), and $\Delta m_{iso}$ (right), with their
fitted corrections for the spiral galaxy M88 in the $K_s$ band.}\label{m88cor}
\end{center}
\end{figure}

The resulting corrections show very comparable trends in the three bands, due to the
similarity of the light profiles of the galaxy in the three bands. To demonstrate the
effectiveness of our procedure we therefore reduce the discussion to the results of the
$K_s$ band only. Figure \ref{m88cor} displays the calculated values of $f(R)$, and
$\Delta m_{iso}$ as a function of $A_{K_s}$, with their fitted relations for the galaxy
M88 in the $K_s$ band. The ability of Eqns. \ref{fit1}, \ref{fit2} to fit the simulated
data varies among the galaxies. Some galaxies show tighter fits compared to M88, while
others show more dispersion. The variation in the quality of the fitting is expected due
to the different structures imprinted on the surface brightness profile of the galaxies.
\subsection{Variation among galaxies (optimized corrections)}\label{sectune}
\begin{figure}
\begin{center}
\includegraphics[width=0.23\textwidth,height=0.23\textwidth]{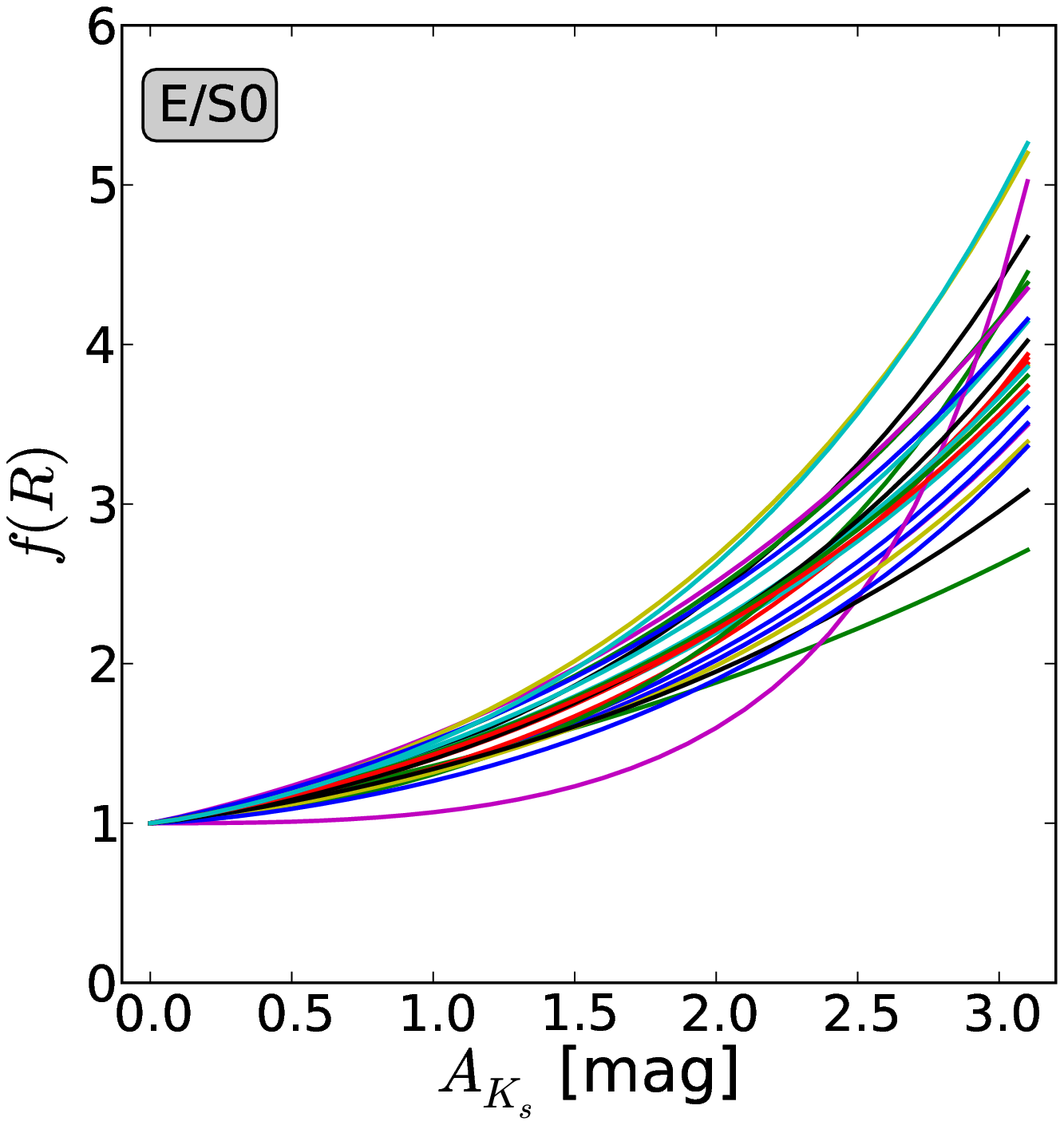}
\includegraphics[width=0.23\textwidth,height=0.23\textwidth]{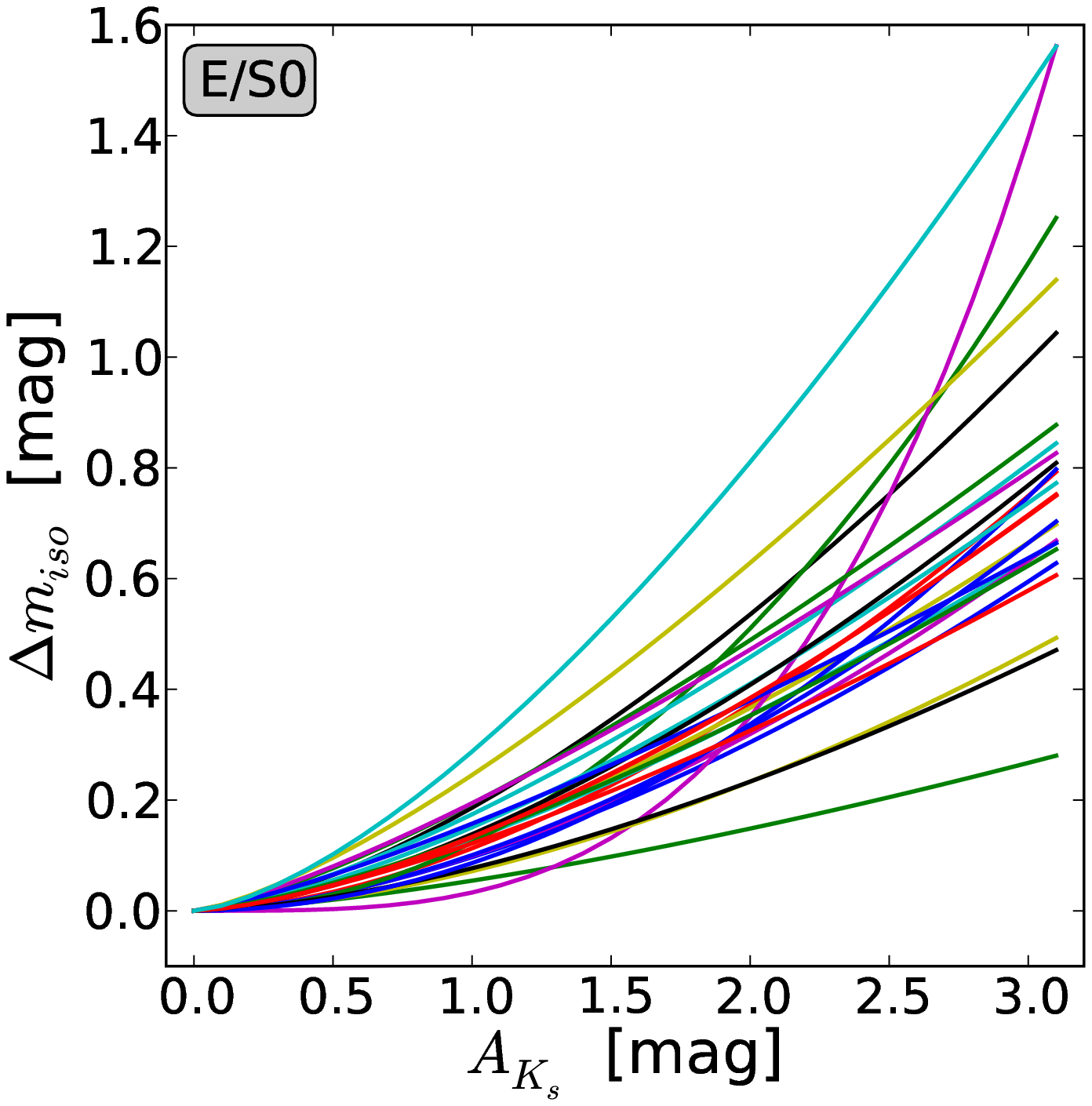}
\includegraphics[width=0.23\textwidth,height=0.23\textwidth]{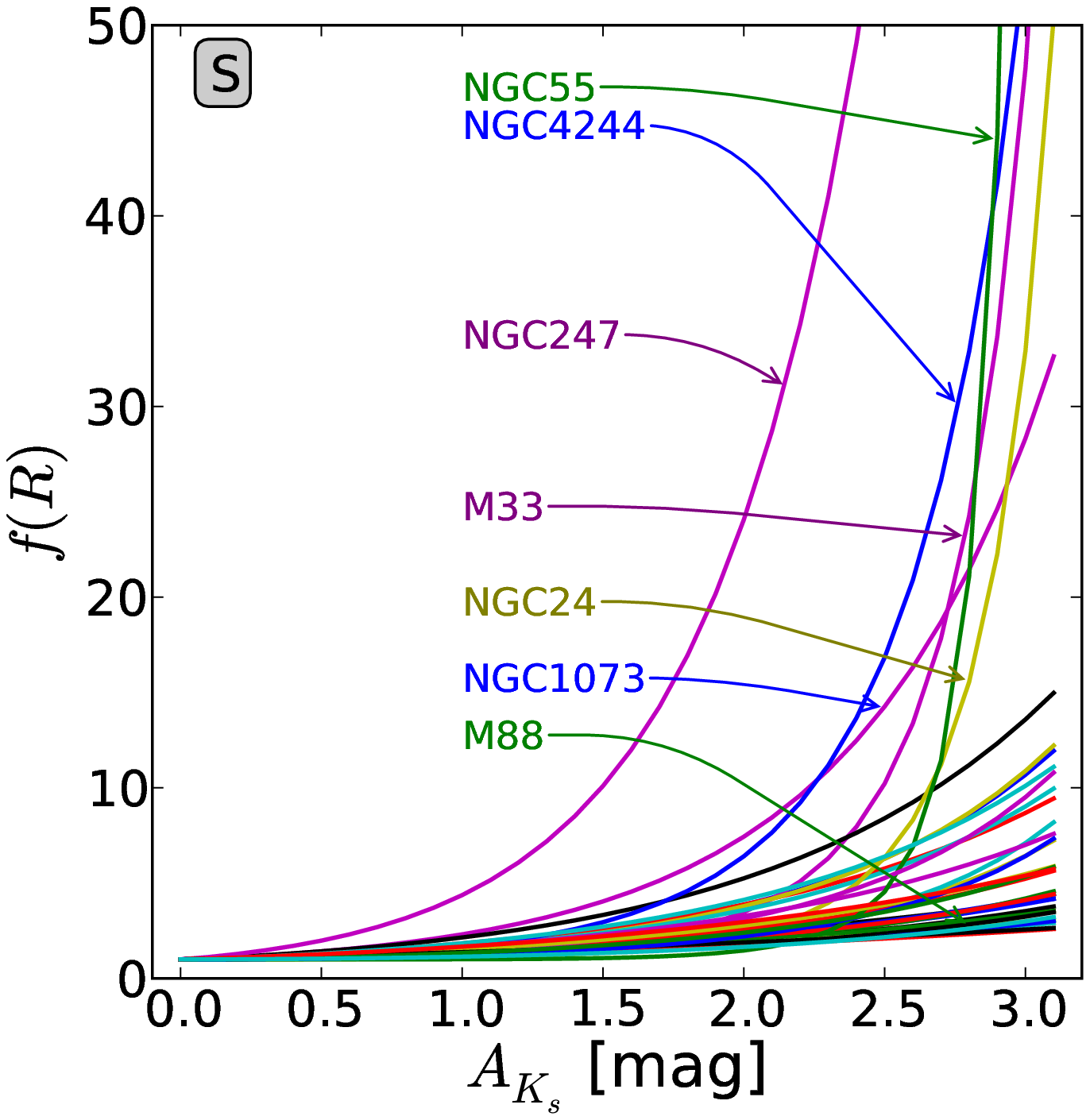}
\includegraphics[width=0.23\textwidth,height=0.23\textwidth]{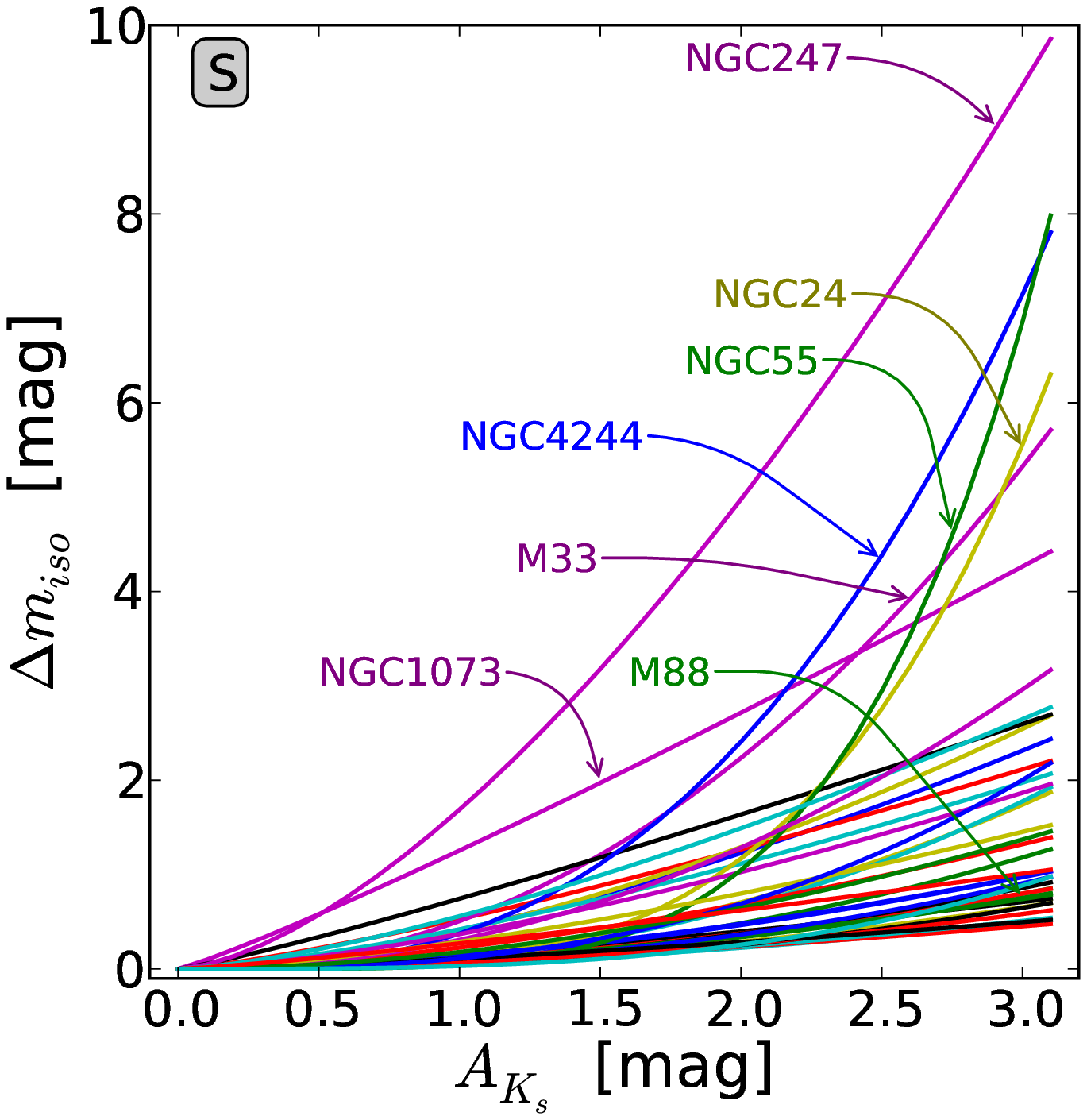}
\caption{Simulated corrections $f(R)$ and $\Delta m_{iso}$. {\bf E/S0} top panel, {\bf S}
bottom panel.}\label{mag-all}
\end{center}
\end{figure}

In Fig. \ref{mag-all} we show the magnitudes and radii corrections for all the {\bf E/~S0}
(top), and {\bf S} galaxies (bottom). The curves in the plots clearly show that the
\citet{cameron} relations in their present form can not fully account for the variation
among the galaxies in the two respective groups.

We investigated the origin of these discrepancies by looking at trends with e.g.
inclination, central surface brightness $\mu_c$, and the disk central surface brightness
$\mu_0$ extrapolated from the fainter outer disk. No significant trend was found between
the corrections and the variation among galaxies inclinations. Most probably a much larger
sample than the current one is needed to investigate the correlation. A clear correlation
between $\mu_0$ and $\mu_c$ with the corrections was noticed, however.

It was noted that galaxies with a low central surface brightness $\mu_c\geq17.0$
mag/arcsec$^2$ in the $K_s$ band, and low surface brightness galaxies in general require
larger corrections. This is expected as those galaxies become obscured much more quickly.
Galaxies that required larger corrections in our sample are M33, NGC24, NGC247, NGC1073,
NGC55, NGC4244; they are labeled in Fig. \ref{mag-all}. Even though the corrections were
found to depend on $\mu_c$, it was in fact found that $\mu_0$ shows the better correlation
with the corrections.

In this section we describe the method to optimize the corrections using either $\mu_0$ or
$\mu_c$. Optimizing the corrections using $\mu_0$ gives tighter estimates to the
corrections than using $\mu_c$. On the other hand, optimizing the corrections
using $\mu_c$ is convenient since it is easily measured and is typically found in large
catalogues including 2MASX. A further point to note is that the NIR $\mu_c$ can be used
to roughly classify the galaxies as early or late-type galaxies \citep{nirmor}. In Sect.
2.3 we derive a more general correction which can be applied when only the galaxies are
classified as early or late type but neither $\mu_0$ nor $\mu_c$ is available.
\subsubsection{Optimized correction based on $\mu_0$}
It was realized that the derived quantity, $\mu_0$, defined as the disk central surface
brightness extrapolated from the fainter outer regions of the galaxy correlates well with
the deviation among the galaxies. The definition of $\mu_0$ corresponds to the disk
central surface brightness for spirals as defined by \citet{freeman}. The better
correlation of the corrections with $\mu_0$ rather than $\mu_c$ is expected since it is
always the fainter outer regions of the galaxy that suffer most from extinction. To
accommodate $\mu_0$ into the corrections, Eqns. \ref{fit1},~\ref{fit2} were re-written as,
\begin{eqnarray}
f(R,\mu_0)&=&10^{a(\mu_0)\cdot(A_\lambda)^{b(\mu_0)}}, \label{fit12}\\
\Delta m_{iso}(\mu_0) &=&F(\mu_0)\cdot(A_\lambda)^{v(\mu_0)}, \label{fit22}
\end{eqnarray}
with
\begin{eqnarray}\label{etune}
a(\mu_0)&=& a_\circ \exp(\mu_0\cdot a_1),\nonumber \\
b(\mu_0)&=& b_\circ \exp(\mu_0\cdot b_1), \\
F(\mu_0)&=& F_\circ \exp(\mu_0\cdot F_1),\nonumber \\
v(\mu_0)&=& v_\circ \exp(\mu_0\cdot v_1).\nonumber
\end{eqnarray}

Using the parameter products $(a\times b)$, $(F\times v)$ for each galaxy, and analyzing
how they vary with $\mu_0$, we derived these parameters
($a_\circ,b_\circ,F_\circ,v_\circ,a_1,b_1,F_1,$ and $v_1$). The values are listed in
Table~\ref{tune}. For each galaxy, the value of $\mu_0$ was found from a linear fitting to
the disk part of the light profile of the galaxy with $\mu_\lambda\gtrsim17.0$
mag/arcsec$^2$. The intercept of the straight line fitted to the disk part of the light
profile with the surface brightness axis was then $\mu_0$. The disk part of the light
profile was identified by visual inspection.
\begin{table}
\begin{center}
\caption{Parameters $a_\circ,\;b_\circ,\;F_\circ,\;v_\circ,\;a_1,\;b_1,\;F_1,$ and $v_1$
for the $\mu_0$ optimization Eqn. \ref{etune}}\label{tune}
\begin{tabular}{lcrrr}
\hline
Galaxy & Param. & $J\;\;\;\;$ & $H\;\;\;$ & $K_s\;\;\;$\\
\hline
&$a_\circ$&0.0005&$5.2(e^{-06})$&0.0002\\
&$a_1$&0.3301&0.6248&0.3993\\
&$b_\circ$&0.3659&15.9697&1.9398\\
{\bf E/S0}&$b_1$&0.0704&$-0.1589$&$-0.0289$\\
&$F_\circ$&$2.8(e^{-11})$&$1.5(e^{-11})$&$6.7(e^{-11})$\\
&$F_1$&1.3090&1.3912&1.3309\\
&$v_\circ$&8.1329&236.9696&18.9573\\
&$v_1$&$-0.0979$&-0.3075&$-0.1561$\\
\\
&$a_\circ$&$3.3(e^{-06})$&$3.9(e^{-06})$&0.0001\\
&$a_1$&0.6093&0.6203&0.4406\\
&$b_\circ$&9.7884&7.0789&2.6909\\
{\bf S}&$b_1$&$-0.1116$&$-0.095$&$-0.0395$\\
&$F_\circ$&$8.1(e^{-11})$&$1.4(e^{-10})$&$2.4(e^{-7})$\\
&$F_1$&1.2277&1.2443&0.8365\\
&$v_\circ$&201.5666&56.9891&17.5403\\
&$v_1$&$-0.2693$&$-0.2055$&$-0.1415$\\
\hline
\end{tabular}
\end{center}
\end{table}

The goodness of $f(R,\mu_0)$ and $\Delta m_{iso}$ in describing the correction curves for
the different galaxies accurately was found to be very sensitive to the value of $\mu_0$.
This shows the importance of having an accurate light profile of the galaxies, and hence
$\mu_0$, to be able to use the $\mu_0$ optimized correction.

\subsubsection{Optimized correction based on $\mu_c$}
For optimizing the extinction corrections using the central surface brightness $\mu_c$ we
used the same methodology as for the $\mu_0$ optimization. To incorporate $\mu_c$ in the
corrections we re-wrote Eqns. \ref{fit1} and \ref{fit2} as,
\begin{eqnarray}
f(R,\mu_c)&=&10^{a(\mu_c)\cdot(A_\lambda)^{b(\mu_c)}}, \label{fitc12}\\
\Delta m_{iso}(\mu_c) &=&F(\mu_c)\cdot(A_\lambda)^{v(\mu_c)}, \label{fitc22}
\end{eqnarray}
with
\begin{eqnarray}\label{cetune}
a(\mu_c)&=& a_\circ \exp(\mu_c\cdot a_1),\nonumber \\
b(\mu_c)&=& b_\circ \exp(\mu_c\cdot b_1), \\
F(\mu_c)&=& F_\circ \exp(\mu_c\cdot F_1),\nonumber \\
v(\mu_c)&=& v_\circ \exp(\mu_c\cdot v_1).\nonumber
\end{eqnarray}
Again we determine the products $(a\times b)$ and $(F\times v)$ and how they vary with
$\mu_c$ and hence derived the parameters ($a_\circ,b_\circ,F_\circ,v_\circ,a_1,b_1,F_1,$
and $v_1$). The values of these parameters are given in Table \ref{ctune}.
\begin{table}
\begin{center}
\caption{Parameters $a_\circ,\;b_\circ,\;F_\circ,\;v_\circ,\;a_1,\;b_1,\;F_1,$ and $v_1$
for the $\mu_c$ optimization Eqn. \ref{cetune}}\label{ctune}
\begin{tabular}{lcrrr}
\hline
Galaxy & Param. & $J\;\;\;\;$ & $H\;\;\;$ & $K_s\;\;\;$\\
\hline
&$a_\circ$&0.1279&0.0547&0.1598\\
&$a_1$&0.0063&0.0707&$-0.0061$\\
&$b_\circ$&0.8397&1.3805&0.7292\\
{\bf E/S0}&$b_1$&0.0248&$-0.0115$&0.0381\\
&$F_\circ$&0.0073&0.0034&0.0151\\
&$F_1$&0.1889&0.2574&0.1632\\
&$v_\circ$&1.4504&2.1299&0.8602\\
&$v_1$&0.0046&$-0.0253$&0.0420\\
\\
&$a_\circ$&0.0068&0.0095&0.0157\\
&$a_1$&0.1953&0.1824&0.1641\\
&$b_\circ$&2.7721&2.5213&1.3653\\
{\bf S}&$b_1$&$-0.0455$&$-0.0397$&0.0009\\
&$F_\circ$&0.0003&0.0007&0.0012\\
&$F_1$&0.4136&0.3833&0.3751\\
&$v_\circ$&6.4910&4.6215&2.3296\\
&$v_1$&$-0.0866$&$-0.0674$&$-0.0267$\\
\hline\end{tabular}
\end{center}
\end{table}

\begin{figure}
\begin{center}
\includegraphics[width=0.23\textwidth,height=0.23\textwidth]{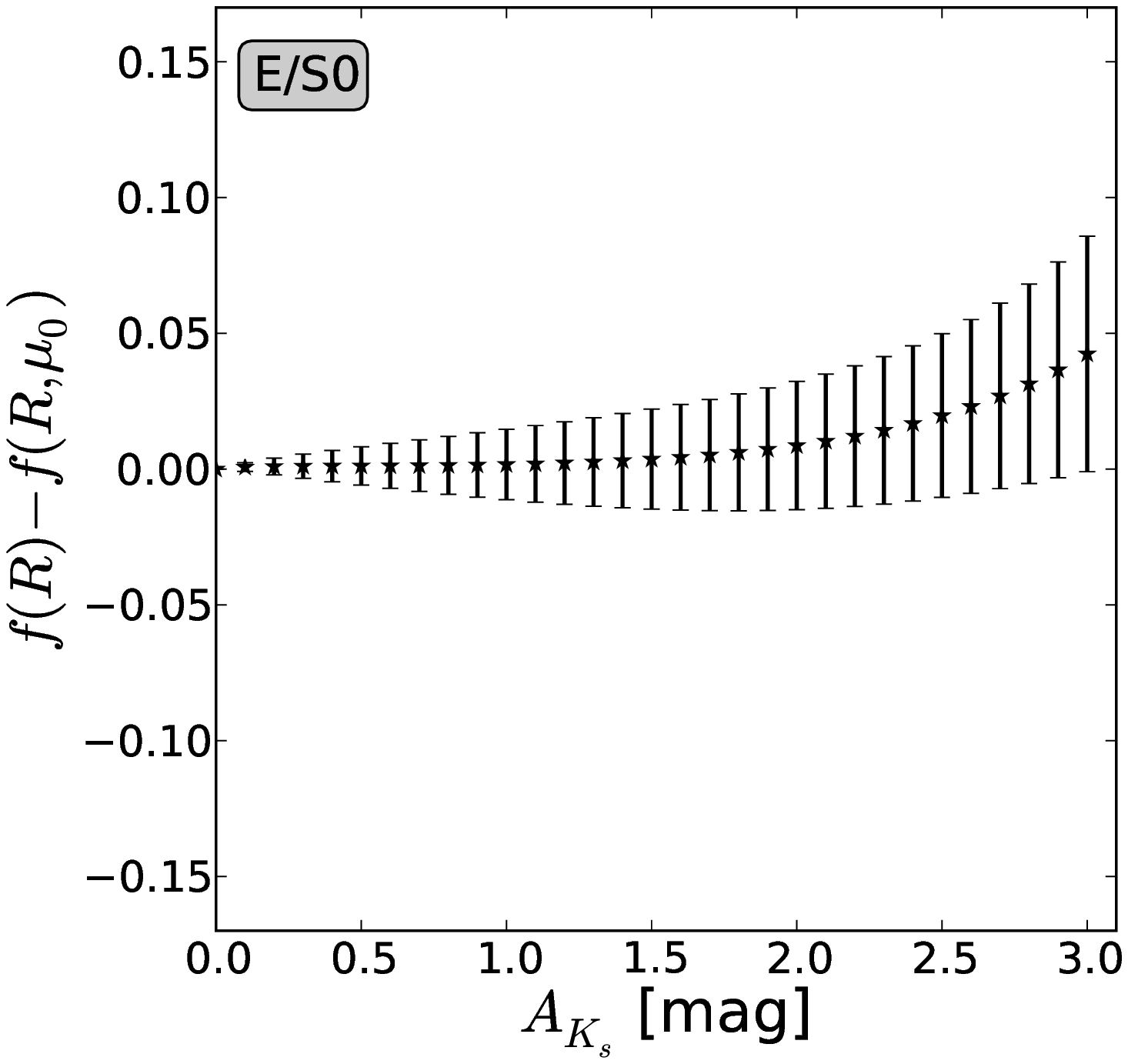}
\includegraphics[width=0.23\textwidth,height=0.23\textwidth]{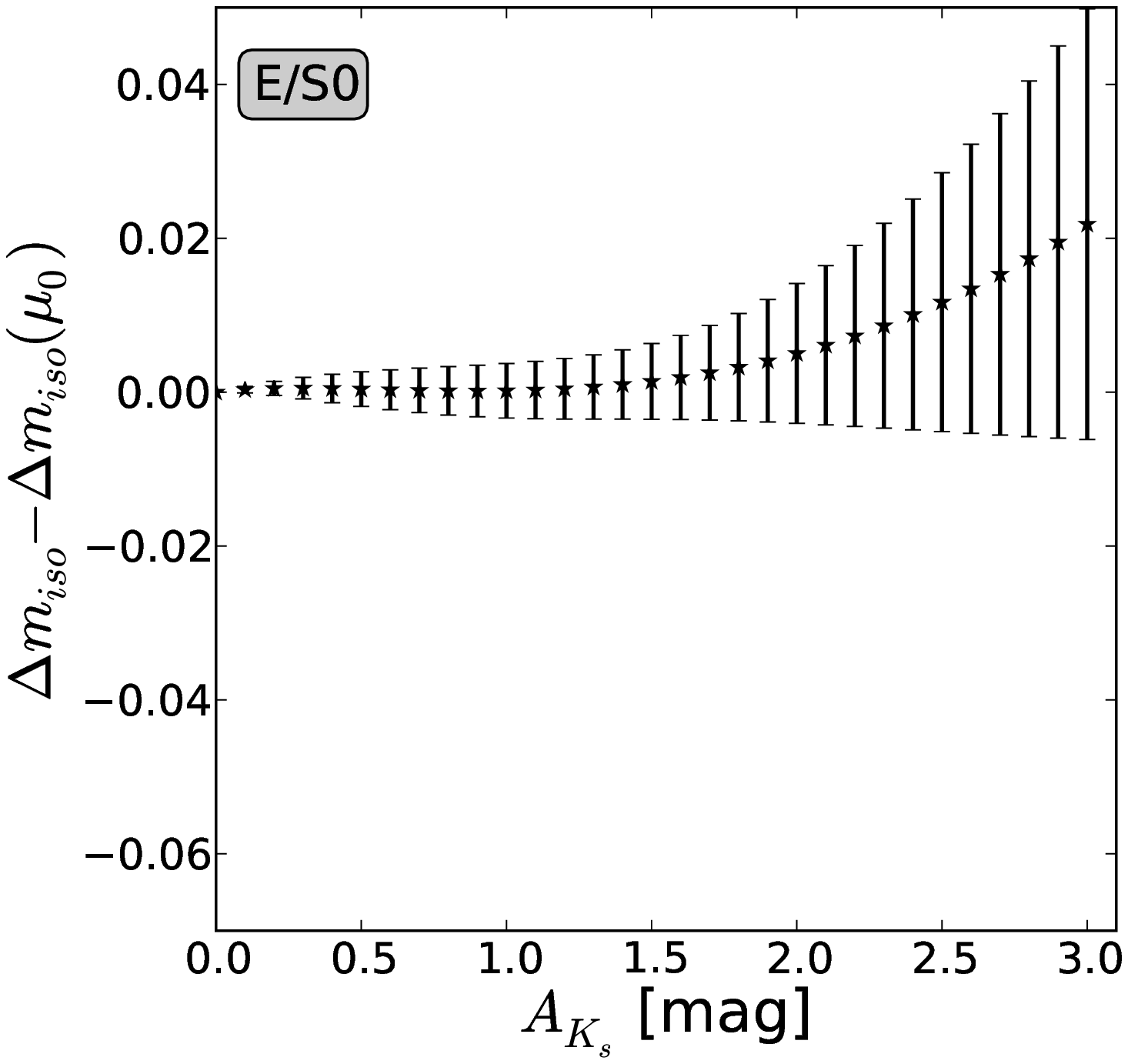}
\includegraphics[width=0.23\textwidth,height=0.23\textwidth]{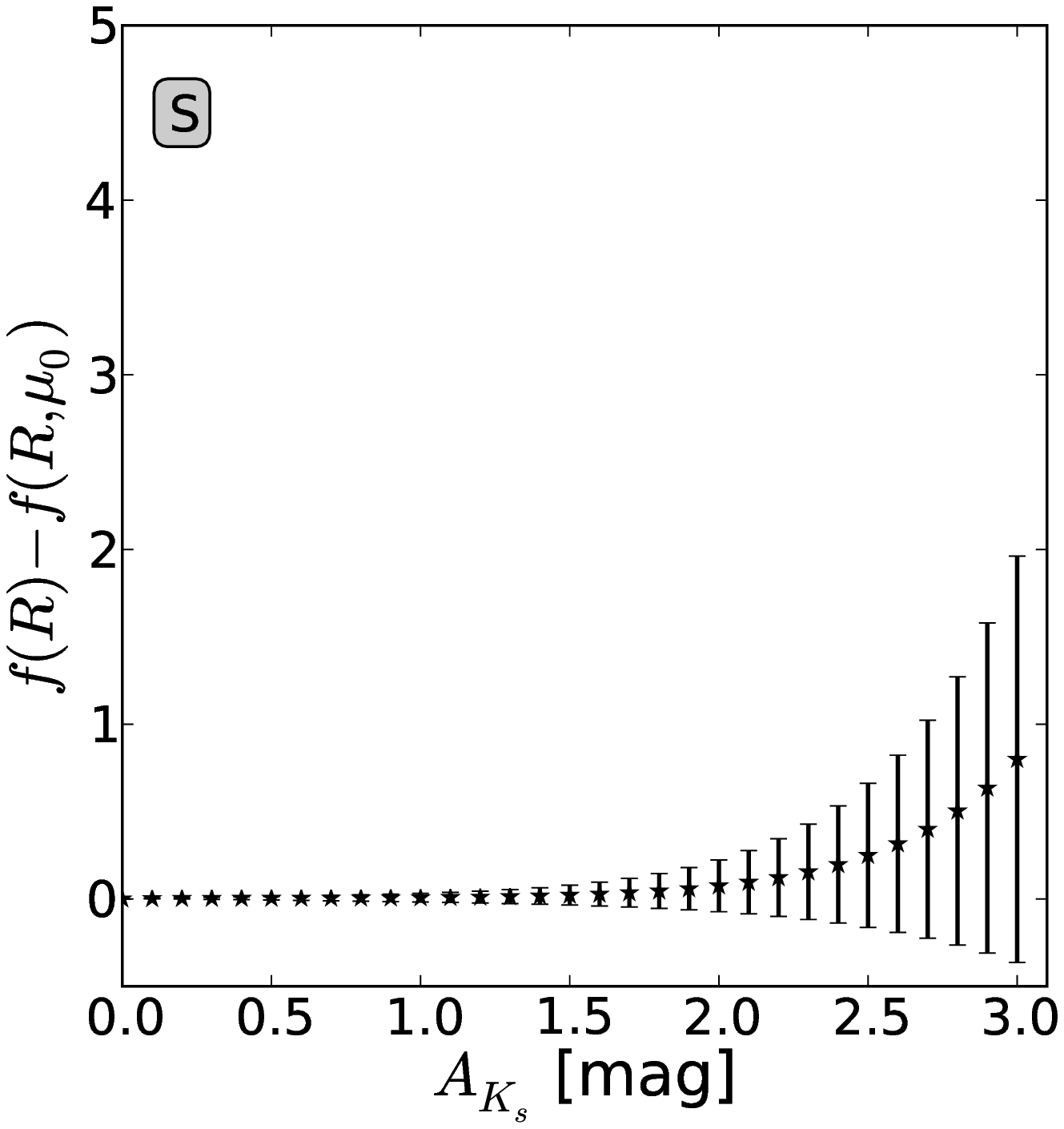}
\includegraphics[width=0.23\textwidth,height=0.23\textwidth]{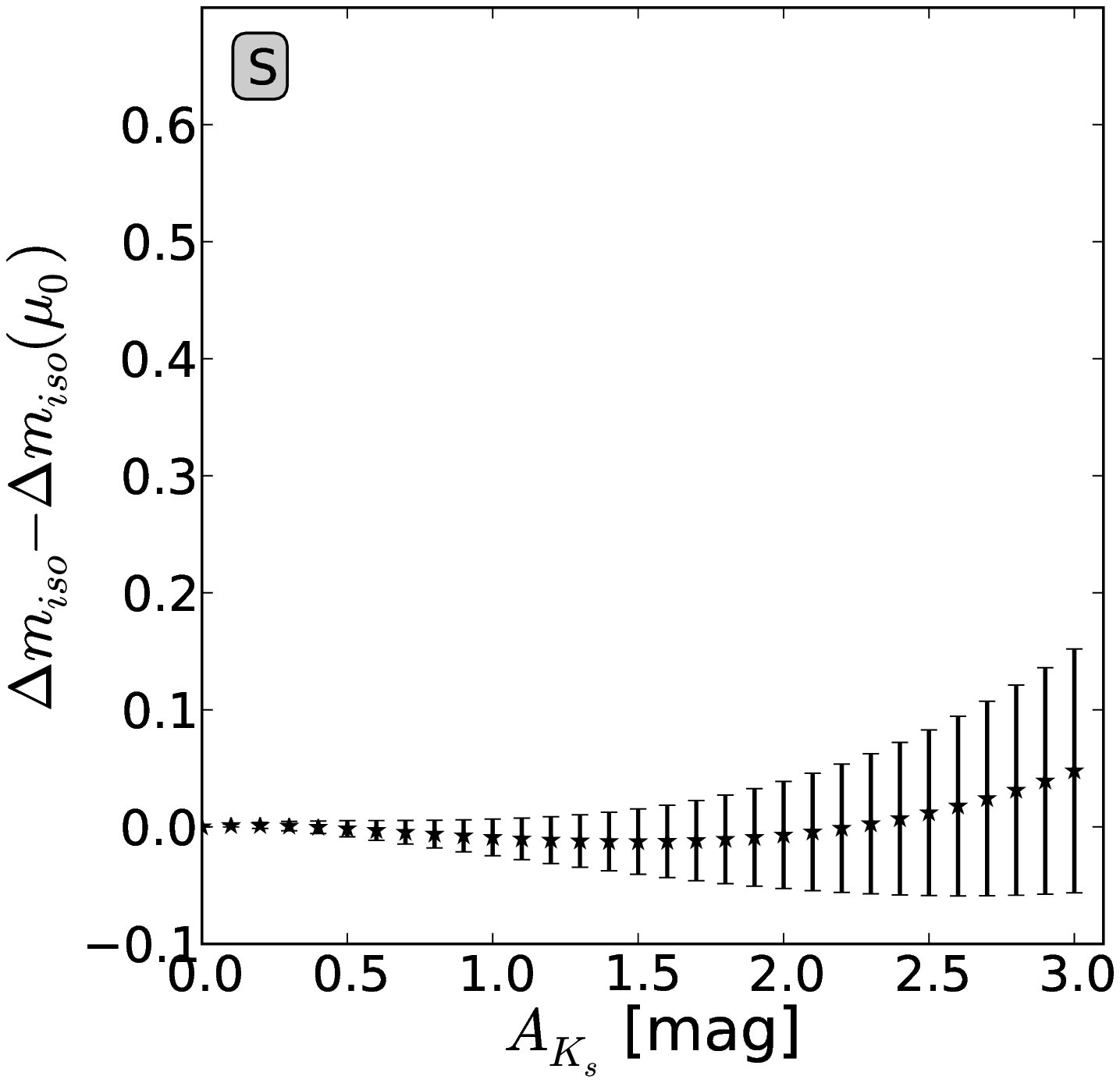}
\caption{Comparison between simulated corrections and $m_0$ optimised corrections derived
using $f(R,\mu_0)$ and $\Delta m_{iso}(R,\mu_0)$, top: {\bf E/S0}, bottom: {S}
galaxies.}\label{tune-comp}
\end{center}
\end{figure}

\begin{figure}
\begin{center}
\includegraphics[width=0.23\textwidth,height=0.23\textwidth]{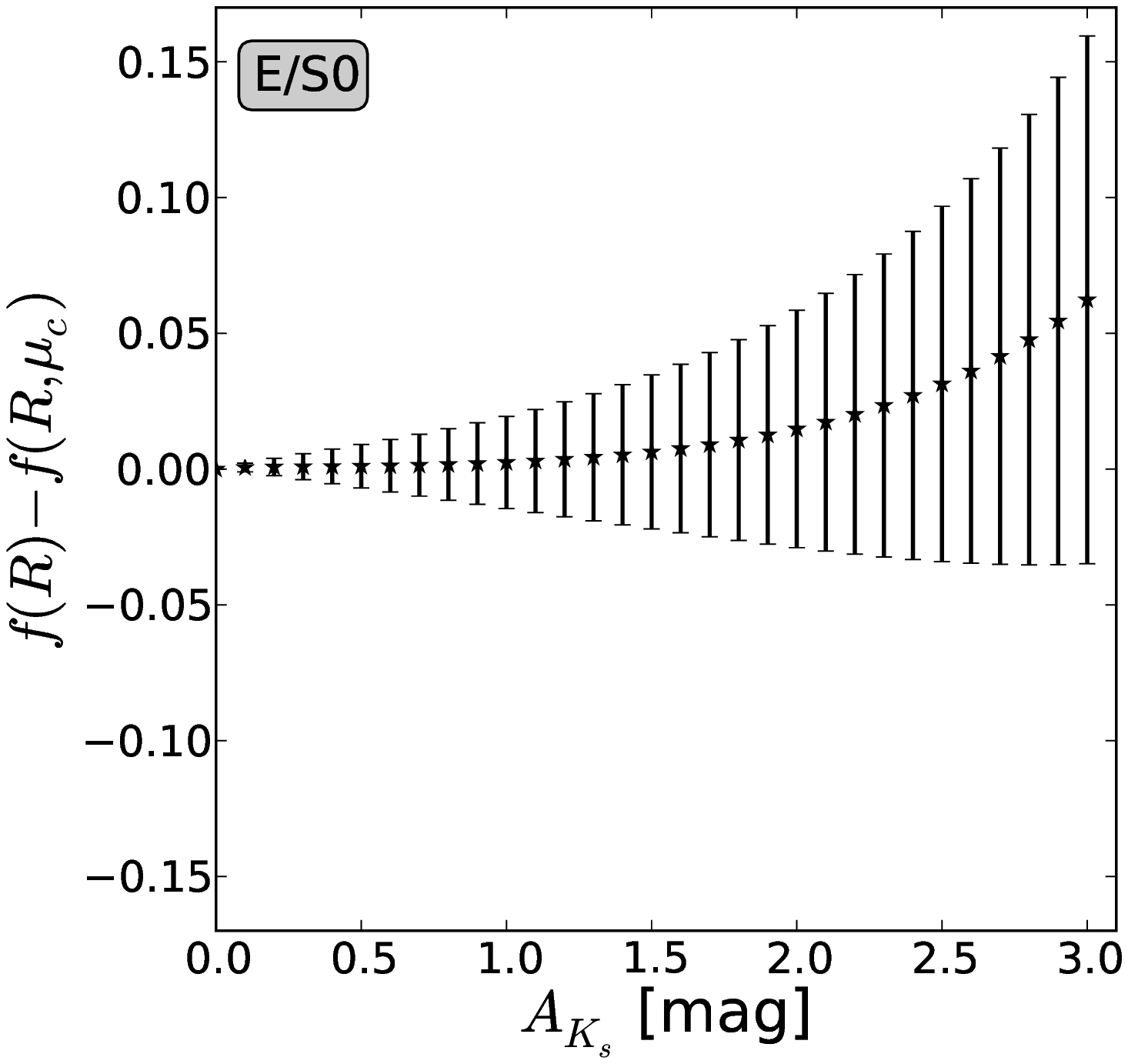}
\includegraphics[width=0.23\textwidth,height=0.23\textwidth]{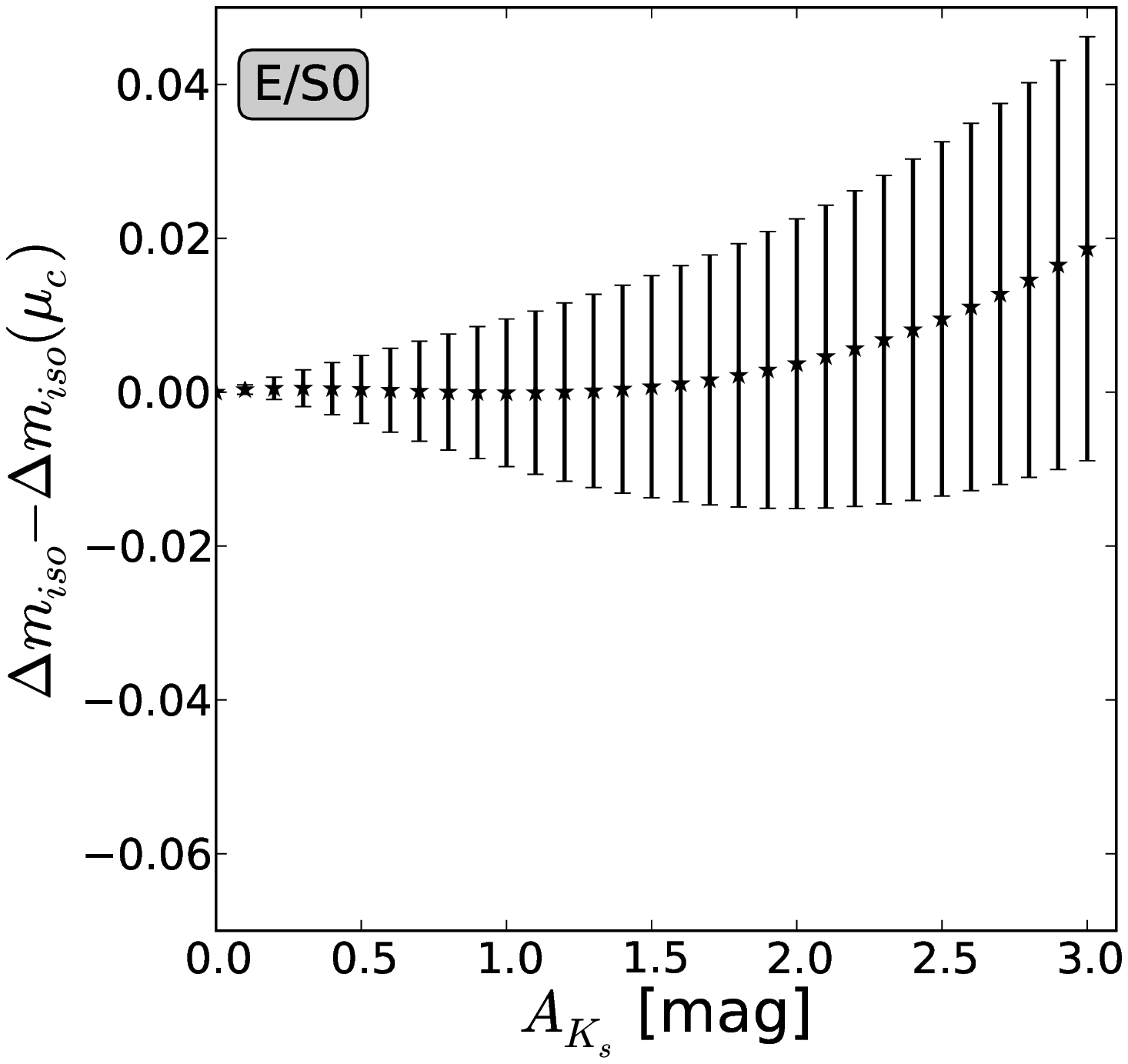}
\includegraphics[width=0.23\textwidth,height=0.23\textwidth]{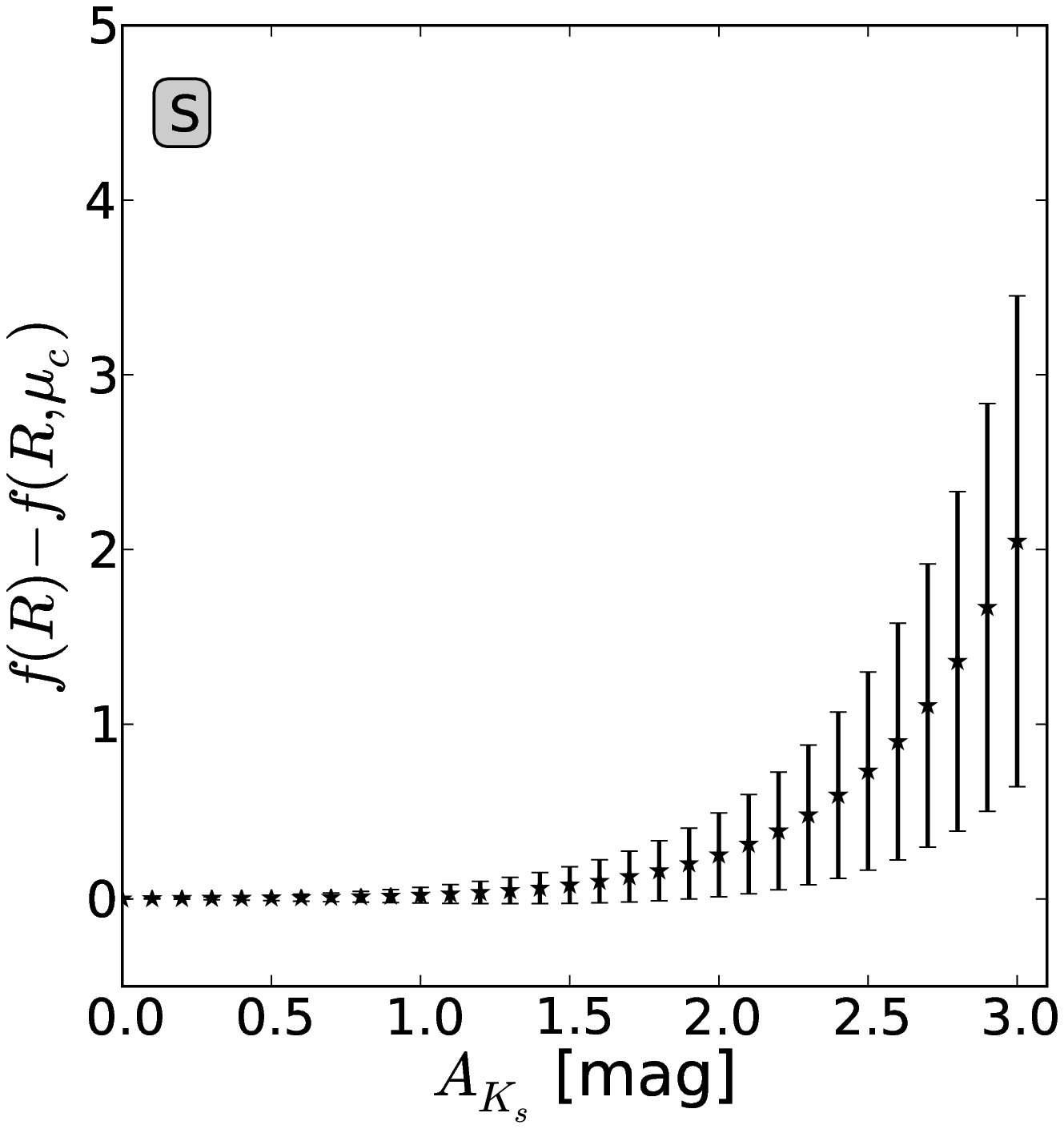}
\includegraphics[width=0.23\textwidth,height=0.23\textwidth]{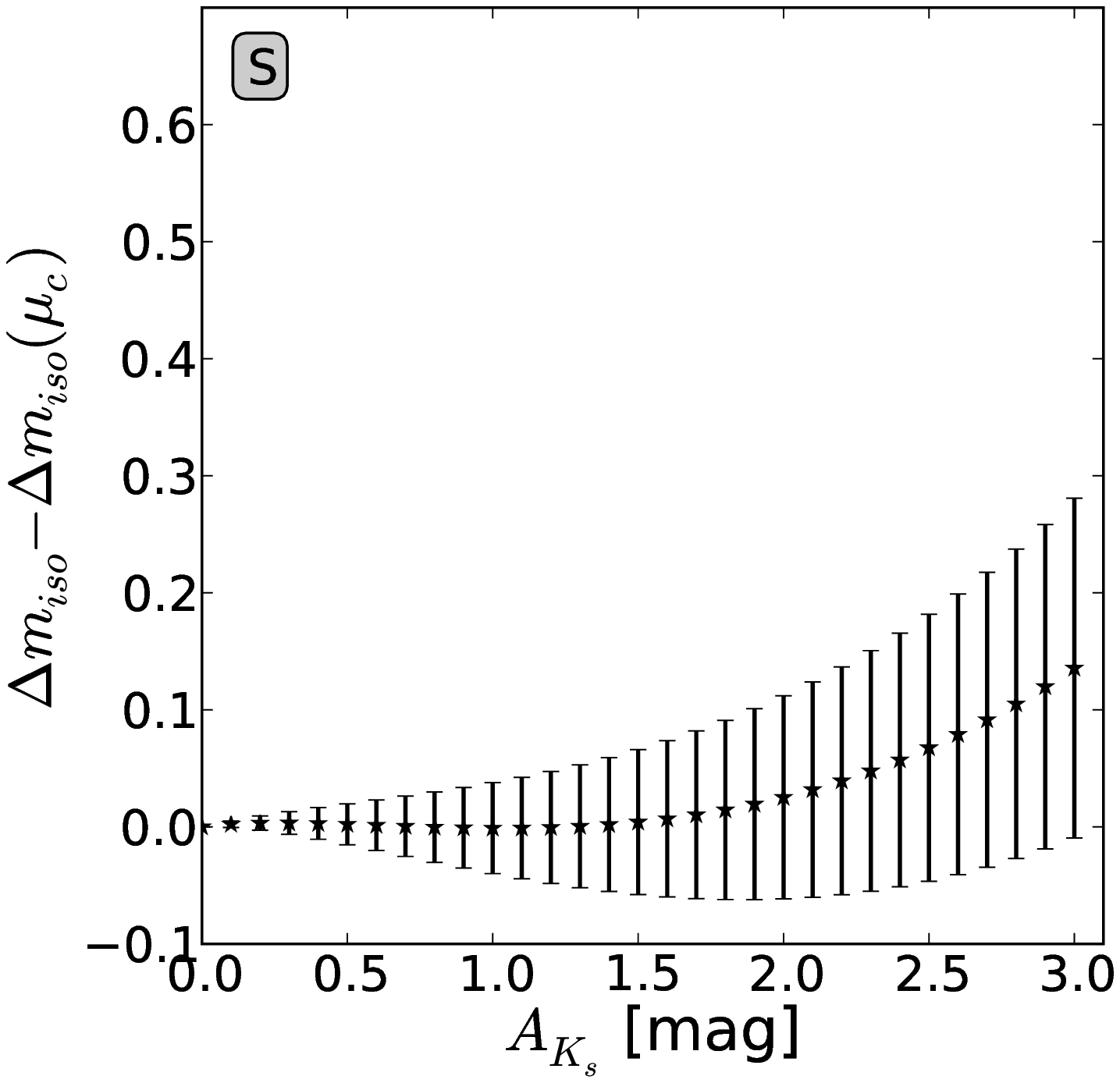}
\caption{Comparison between simulated corrections and the $\mu_c$ optimised corrections
derived using $f(R,\mu_c)$ and $\Delta m_{iso}(R,\mu_c)$, top: {\bf E/S0}, bottom: {S}
galaxies.}\label{ctune-comp}
\end{center}
\end{figure}

To quantify the performance of the two optimization corrections, we computed the
difference between the corrections as given by the simulated data and as given by the
optimized corrections using both the parametrization $f(R,\mu_0)$, $\Delta m_{iso}(\mu_0)$
and using $f(R,\mu_c)$, $\Delta m_{iso}(\mu_c)$. The comparison was made for all galaxies
in the $K_s$ band in the extinction range $A_{K_s}=0\fm0-3\fm0$. A summary of the
comparisons is listed in Table \ref{ttune}, and displayed in Fig. \ref{tune-comp} for the
$\mu_0$ optimization, and Table \ref{tctune}, Fig. \ref{ctune-comp} for the $\mu_c$
optimization.
\begin{table*}
\begin{minipage}{130mm}
\begin{center}
\caption{Comparison between simulated corrections and $\mu_0$ optimized corrections at the
extinction levels $A_{K_s}=~0\fm5$, $1\fm0$, $2\fm0$ and $3\fm0$.}\label{ttune}
\begin{tabular}{llrrrrrrrr}
\hline
Galaxy&Param.&$0\fm5\;\;$&$SDE\;$&$1\fm0\;\;$&$SDE\;$&$2\fm0\;\;$&$SDE\;$&$3\fm0 \;\;$&$SD
E\;$\\
\hline
{\bf E/S0} &$f(R)$&0.0012&0.0071&0.0017&0.0130&0.0087&0.0236&0.0424&0.0433\\
{\bf E/S0} &$\Delta m_{iso}$&0.0004&0.0022&0.0002&0.0035&0.0050&0.0091&0.0218&0.0280\\
{\bf S}&$f(R)$&0.0042&0.0104&0.0071&0.0248&0.0748&0.1476&0.7993&1.1632\\
{\bf S}&$\Delta
m_{iso}$&$-0.0016$&0.0069&$-0.0090$&0.0156&$-0.0070$&0.0457&0.0479&0.1042\\
\hline
\end{tabular}
\end{center}
The $\mu_0$ optimized corrections were calculated using Eqns. \ref{fit12}, \ref{fit22}.
The value of $\mu_0$ for each galaxy was found by performing a linear regression to
fainter outer region of the galaxy light profile with $\mu_\lambda\gtrsim17.0$
mag/arcsec$^2$ and measuring its intercept with the ordinate.
\end{minipage}
\end{table*}
\begin{table*}
\begin{minipage}{130mm}
\begin{center}
\caption{Comparison between simulated corrections and $\mu_c$ optimized corrections at the
extinction levels $A_{K_s}=~0\fm5$, $1\fm0$, $2\fm0$ and $3\fm0$.}\label{tctune}
\begin{tabular}{llrrrrrrrr}
\hline
Galaxy&Param.&$0\fm5\;\;$&$SDE\;$&$1\fm0\;\;$&$SDE\;$&$2\fm0\;\;$&$SDE\;$&$3\fm0 \;\;$&$SD
E\;$\\
\hline
{\bf E/S0}&$f(R)$&0.0010&0.0080&0.0025&0.0170&0.0148&0.0437&0.0623&0.0972\\
{\bf E/S0}&$\Delta m_{iso}$&0.0004&0.0044&0.0000&0.0096&0.0037&0.0188&0.0186&0.0276\\
{\bf S}&$f(R)$&0.0057&0.0157&0.0215&0.0449&0.2520&0.2398&2.0472&1.4042\\
{\bf S}&$\Delta m_{iso}$&0.0022&0.0175&$-0.0011$&0.0388&0.0252&0.0868&0.1356&0.1451\\
\hline
\end{tabular}
\end{center}
The $\mu_c$ optimized corrections were calculated using Eqns. \ref{fitc12}, \ref{fitc22}.
\end{minipage}
\end{table*}

The error bars in Figs. \ref{tune-comp}, \ref{ctune-comp} and  the plots that follow are
the standard deviation error defined as $SDE=\sigma/\sqrt{N}$, with $\sigma$, the standard
deviation of the binned values and $N$ the number of data points in the bin, which are 25
for {\bf E/S0}, and 39 for {\bf S}.

The plots and tables show that the accuracy of the corrections are extinction dependent,
showing smaller deviations and less dispersion at low levels of extinction. It is also
evident that spiral galaxies show larger deviations and more dispersion than elliptical
and lenticular galaxies. The relatively larger errors for the spiral galaxies can be
attributed to the more structural features like spiral arms, bars traced by their light
profiles. The comparison also shows that the corrections parametrized using the disk
central surface brightness show less systematic shifts and tighter dispersion than the
corrections using the central surface brightness optimization. The better performace of
$\mu_c$ in describing the corrections results from the fact that it is the outer galaxy
disk that suffers more of the obscuration.

It is also worth noting that $\approx 85\%,\; 94\%$ and $98\%$ of the galaxies in the
$J,\; H$ and $K_s$ bands in our ZOA survey are in regions with an extinction
$A_{J,H,K_s}\lesssim1\fm0$ where the corrections show very little dispersion.
\subsection{Average behaviour (average correction)}\label{secavg}
The application of the optimized method to estimate the obscuration corrections for
galaxies requires the knowledge of the value of $\mu_0$ and hence the galaxy light
profile. In many cases that information is not available, or hard to obtain, like trying
to correct isophotal radii or magnitudes for galaxies in a galaxy catalogue such as the
2MASX. For such cases an average correction independent of the galaxy light profile is
required. In this section we derive such an average correction. A comparison between the
application of the optimized and average corrections is given later in Sect. 3.

\begin{figure}
\begin{center}
\includegraphics[width=0.23\textwidth,height=0.23\textwidth]{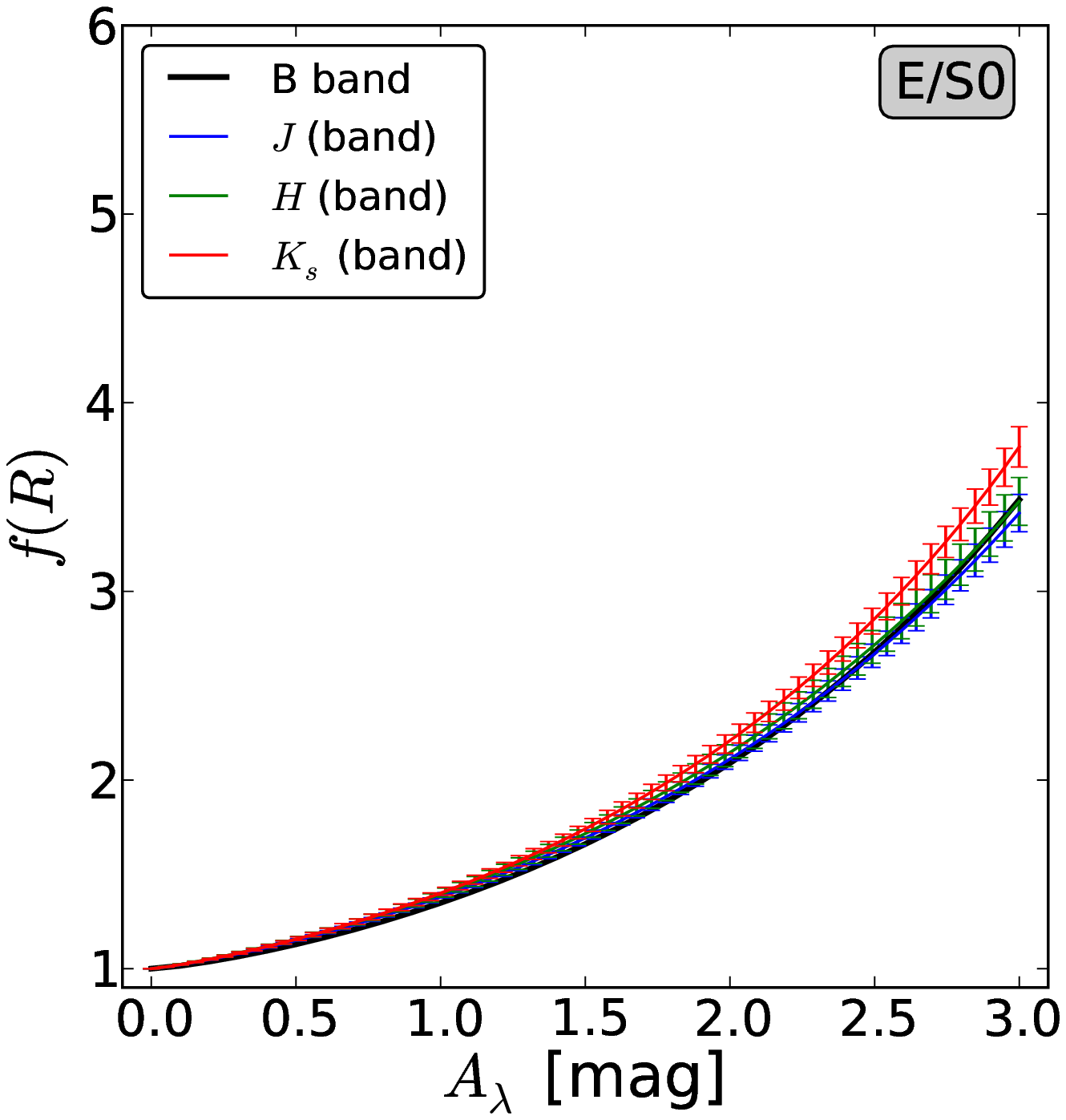}
\includegraphics[width=0.23\textwidth,height=0.23\textwidth]{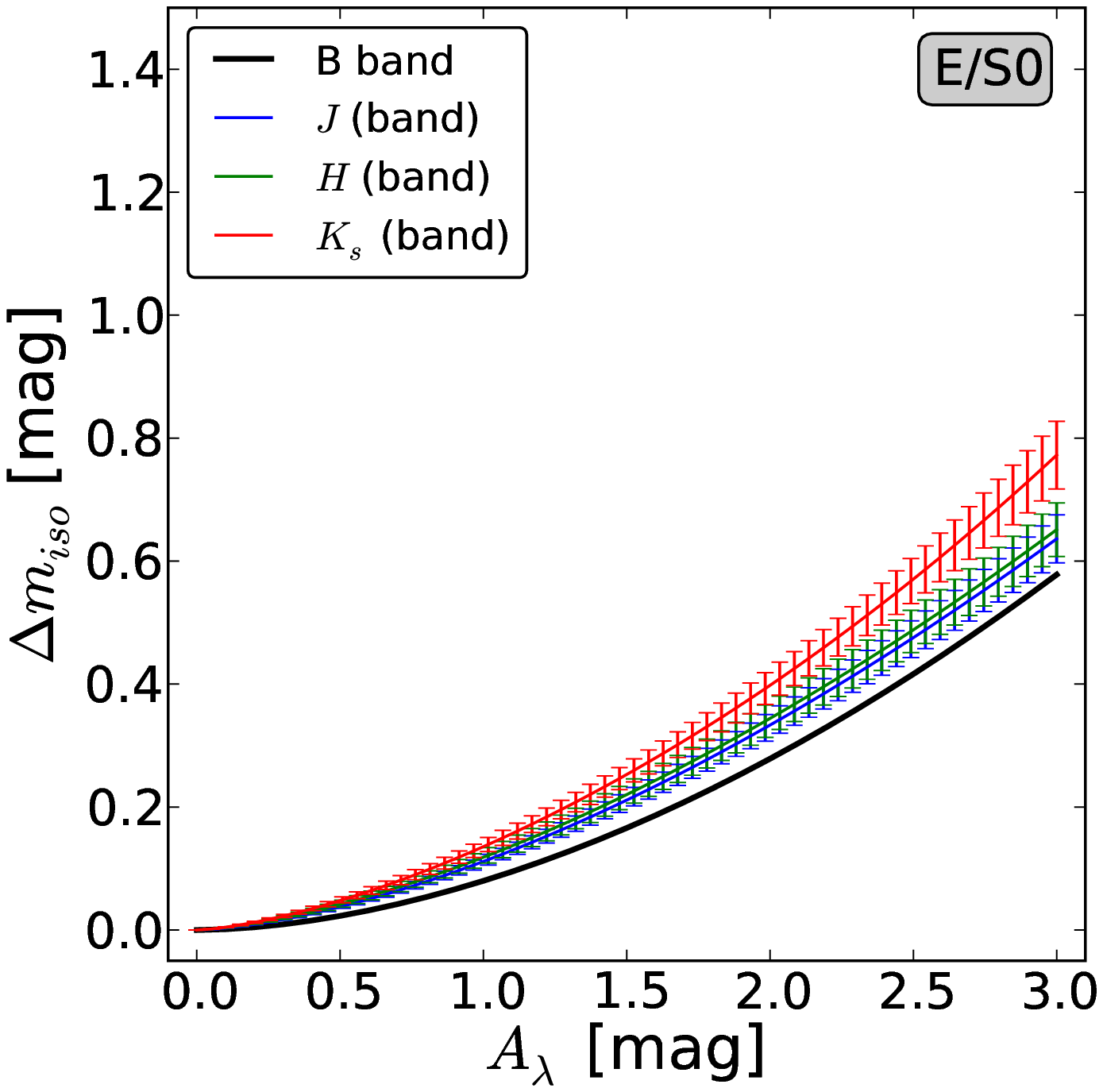}
\includegraphics[width=0.23\textwidth,height=0.23\textwidth]{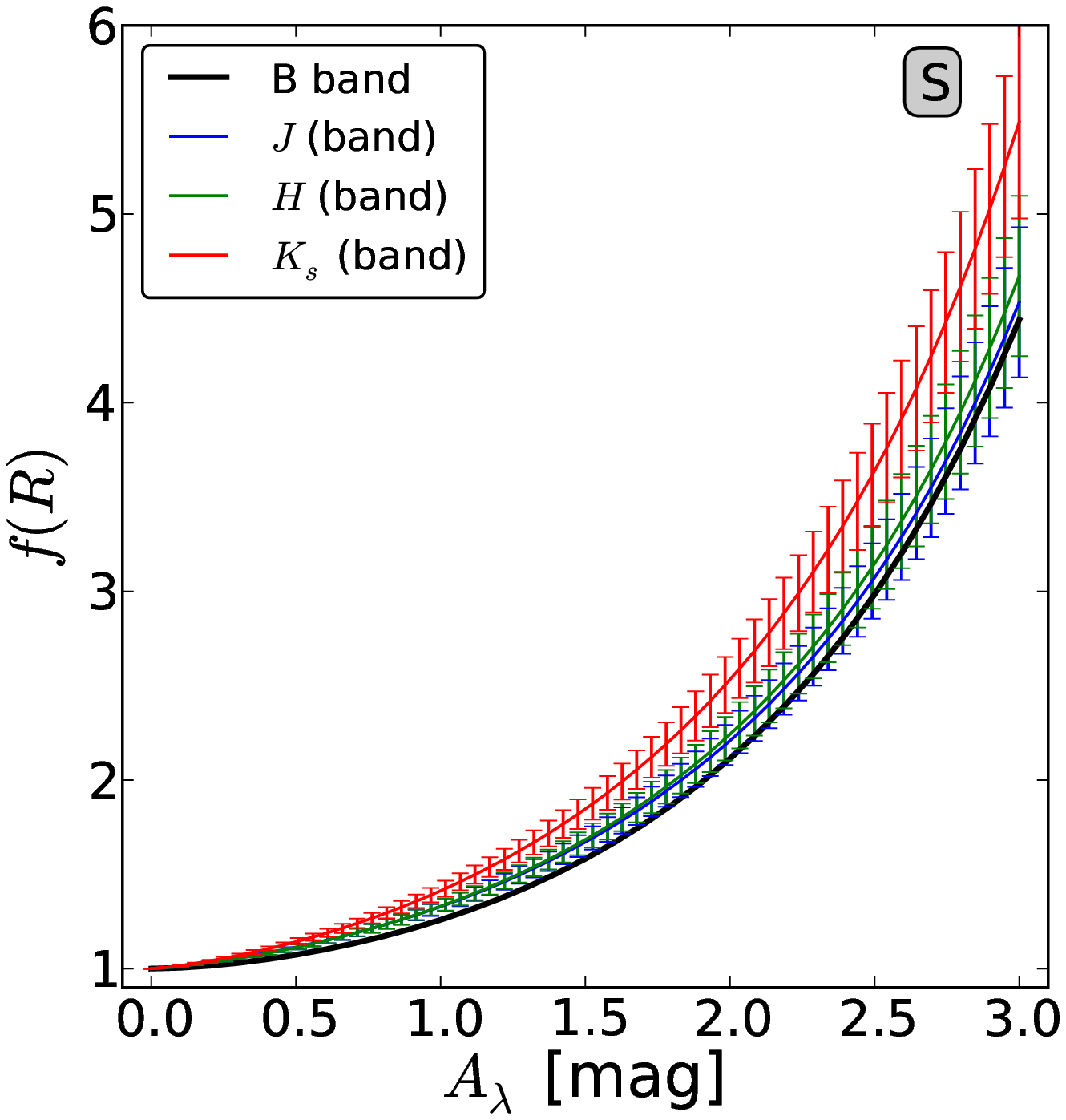}
\includegraphics[width=0.23\textwidth,height=0.23\textwidth]{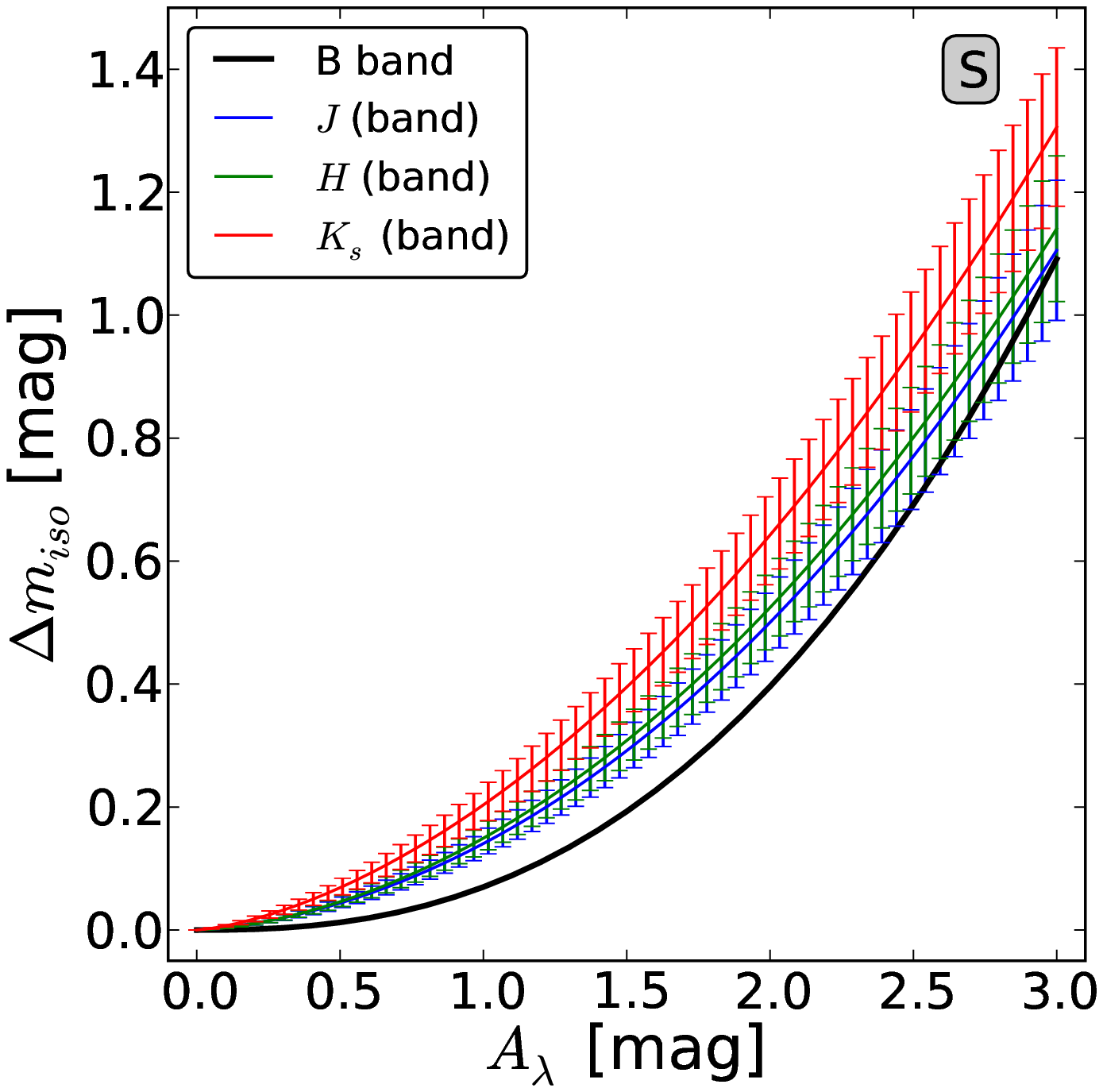}
\caption{Average correction curves for the 25 {\bf E/S0} (top), and 33 {\bf S} galaxies
(bottom) representing the average correction in each respective galaxy class, for the
bands $J$, $H$, $K_s$ and $B$ band.} \label{avg-plots}
\end{center}
\end{figure}

The all elliptical galaxies and the majority of the spiral galaxies ($85\%$ of the
spirals) show comparable trends (see Fig. \ref{mag-all}), while only a few galaxies (M33,
NGC24, NGC247, NGC1073, NGC55, NGC4244, marked on the plots) deviate from the average
behaviour. To produce an average correction for each galaxy family {\bf E/S0} or {\bf S},
we excluded the strongly deviating galaxies. For the ones showing the similar behaviour
(25 {\bf E/S0} and 33 {\bf S} galaxies) their simulated corrections were binned, resulting
in an average correction curve for each family in each of the three bands. These average
curves for the $J$, $H$ and $K_s$ bands are displayed in Fig. \ref{avg-plots}. For
comparison the $B$-band corrections are displayed as well. The latter are taken from
\cite{cameron}.

As expected, the error bars grow with increasing extinction. It should be noted that the
error bars are biased with our selection of galaxies representing the average behaviour.
If we would include the strongly deviating galaxies, the error bars would obviously be
larger. But given that the majority of the galaxies follow the general behaviour it would
be unreasonable to include those outliers for the correction. It is also obvious from Fig.
\ref{avg-plots} that our correction curves are similar to those given by Cameron for the
$B$ band, especially the shorter $J$ and $H$ bands. The average correction curves were
then fitted with Eqns. \ref{fit1}, \ref{fit2}. The respective fitting parameters are given
in Table \ref{tavgfit}.
\setlength{\tabcolsep}{2pt}
\begin{table}
\begin{center}
\caption{Average fitting parameters $a,\:b,\:F,\textrm{and} \:v$ for {\bf E/S0} and {\bf
S} galaxies.}\label{tavgfit}
\begin{tabular}{lcccc}
\hline
Galaxy & Param. & $J$ & $H$ & $K_s$ \\
\hline
&a&$0.1393\pm0.0001$&$0.1438\pm0.0001$&$0.1445\pm0.0003$\\
{\bf E/S0}&b&$1.2212\pm0.0005$&$1.2056\pm0.0004$&$1.2555\pm0.0018$\\
&F&$0.1111\pm0.0001$&$0.1175\pm0.0002$&$0.1328\pm0.0005$\\
&v&$1.5872\pm0.0008$&$1.5559\pm0.0019$&$1.5950\pm0.0041$\\
\\
&a&$0.1169\pm0.0007$&$0.1185\pm0.0006$&$0.1453\pm0.0005$\\
{\bf S}&b&$1.5650\pm0.0056$&$1.5718\pm0.0047$&$1.4778\pm0.0032$\\
&F&$0.1346\pm0.0008$&$0.1442\pm0.0007$&$0.1982\pm0.0009$\\
&v&$1.9080\pm0.0064$&$1.8755\pm0.0053$&$1.7093\pm0.0049$\\
\hline
\end{tabular}
\end{center}
\end{table}

\begin{figure}
\begin{center}
\includegraphics[width=0.23\textwidth,height=0.21\textwidth]{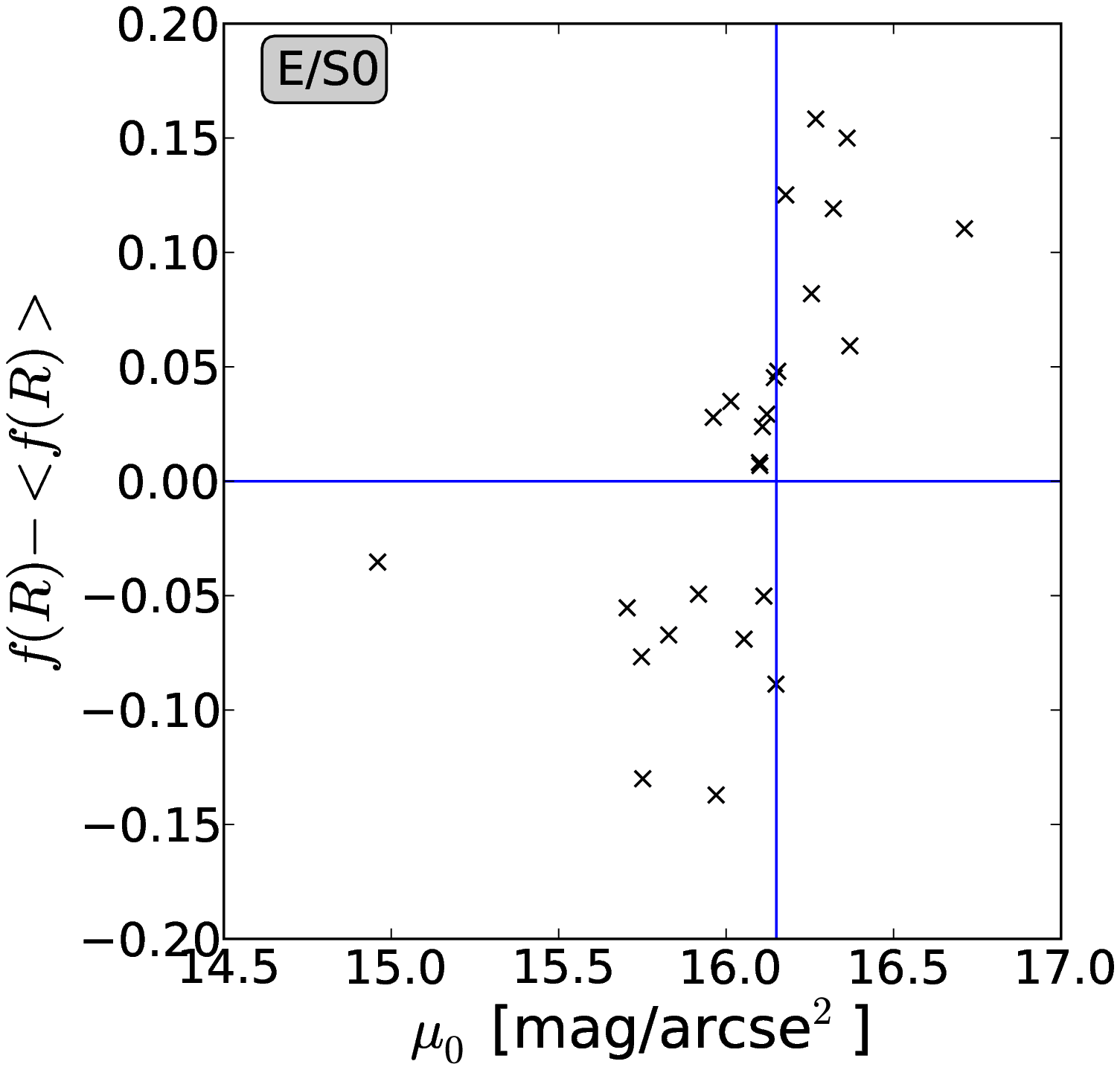}
\includegraphics[width=0.23\textwidth,height=0.21\textwidth]{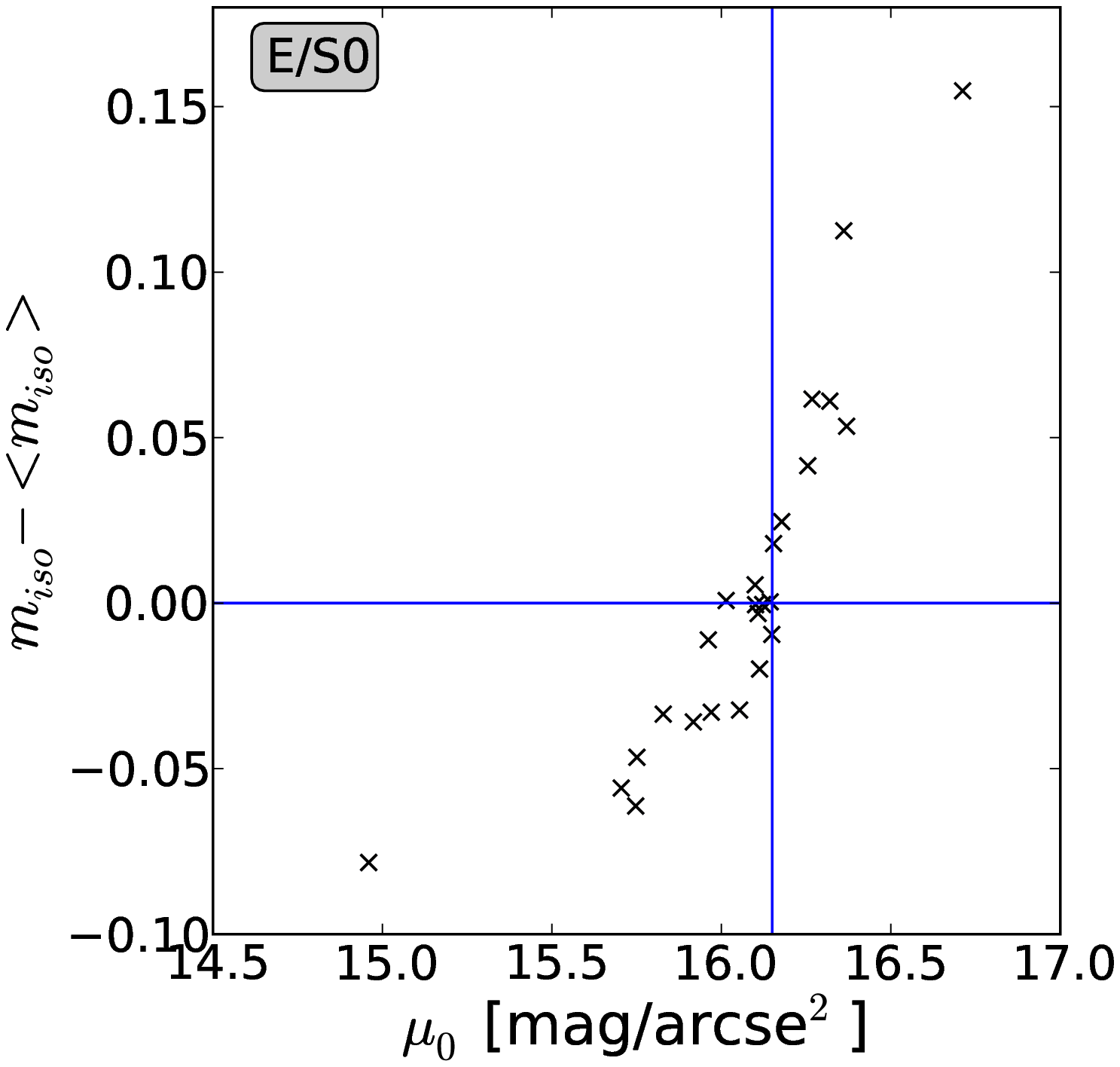}
\includegraphics[width=0.23\textwidth,height=0.21\textwidth]{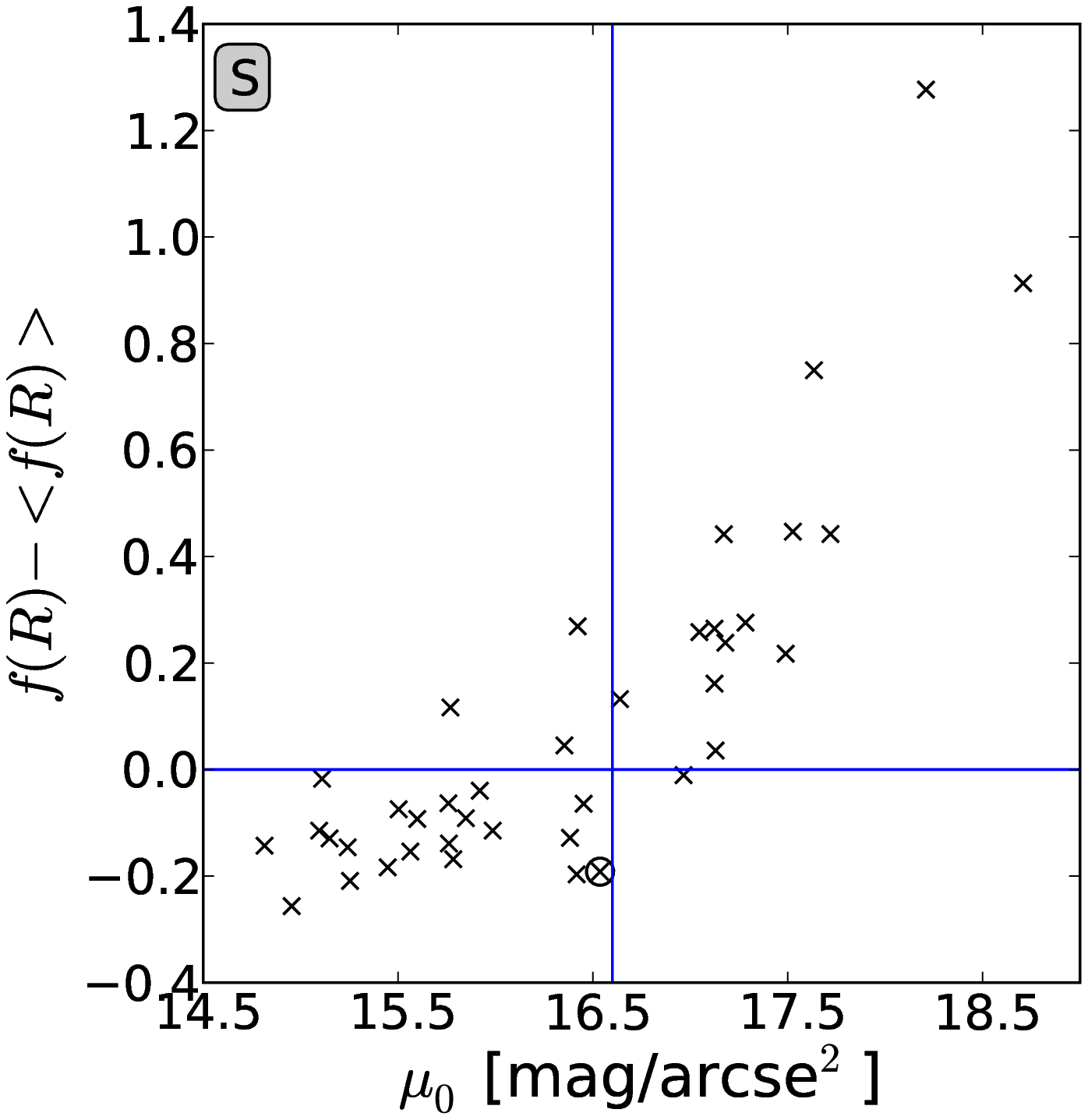}
\includegraphics[width=0.23\textwidth,height=0.21\textwidth]{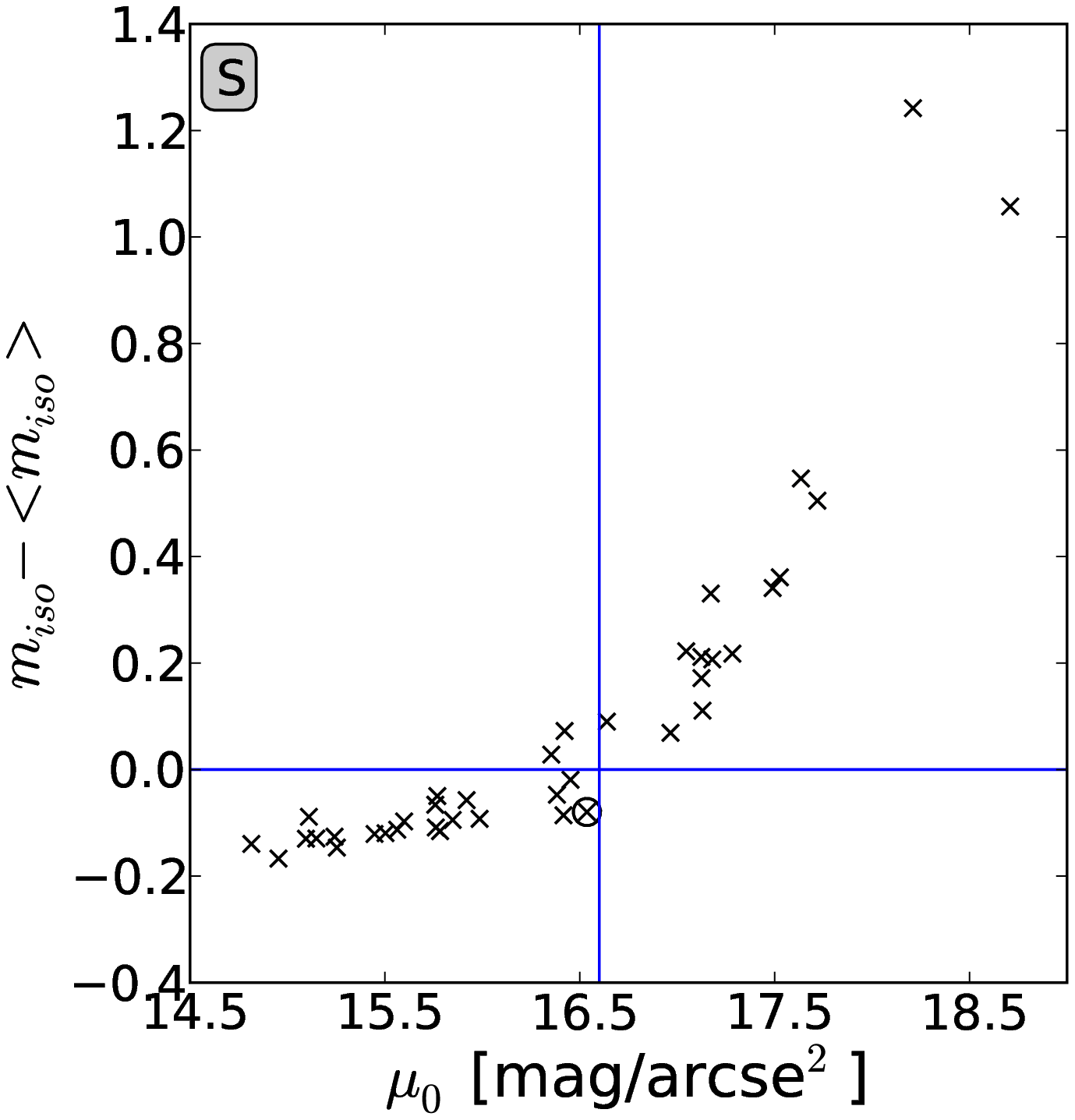}
\caption{Difference between simulated corrections and average correction for an extinction
value of $A_{K_s}=1\fm0$. Top: {\bf E/S0}, radius correction left panel, magnitude
correction right panel. Bottom: {\bf S}, radius correction left panel, magnitude
correction right panel. The values for the galaxy M88 are indicated on the spiral panels
with the symbol $\otimes$.}\label{diff}
\end{center}
\end{figure}

To gain more insight on the variation of the average correction among the galaxies, we
calculated the difference between the simulated corrections and the average corrections
for each galaxy at $A_{K_s}=1\fm0$. The average correction was calculated using Eqns.
\ref{fit1}, \ref{fit2} and the parameters in Table \ref{tavgfit}. In Fig. \ref{diff} we
plot these differences against the value of $\mu_0$ for the galaxy in the $K_s$ band. We
find that {\bf E/S0} galaxies with $\mu_0\geq16.2$ mag/arcsec$^2$ and {\bf S} galaxies
with $\mu_0\geq16.6$ mag/arcsec$^2$ are generally underestimated by the average
correction, while brighter galaxies are overestimated by the correction.

\begin{figure}
\begin{center}
\includegraphics[width=0.22\textwidth,height=0.22\textwidth]{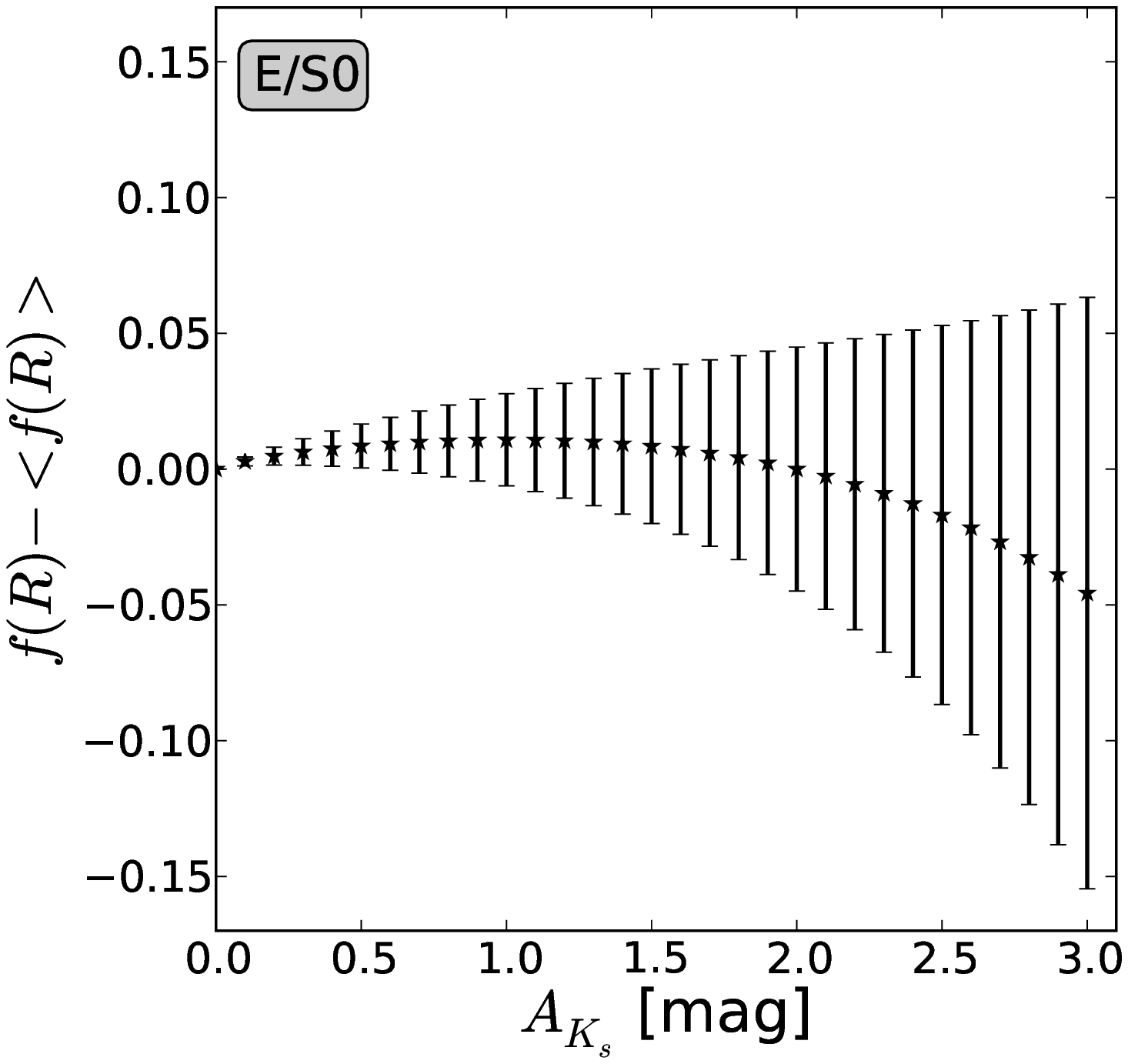}
\includegraphics[width=0.22\textwidth,height=0.22\textwidth]{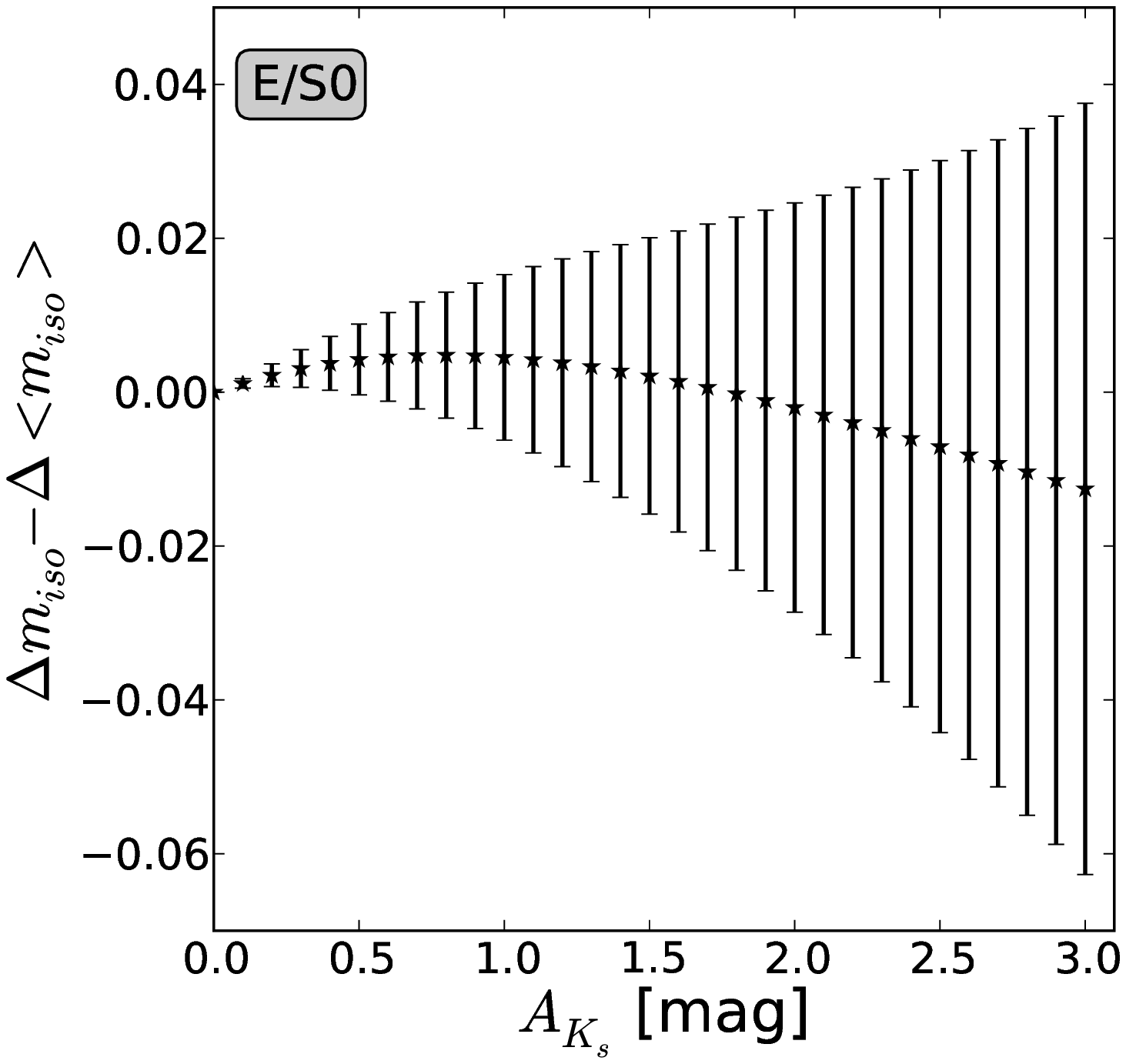}
\includegraphics[width=0.22\textwidth,height=0.22\textwidth]{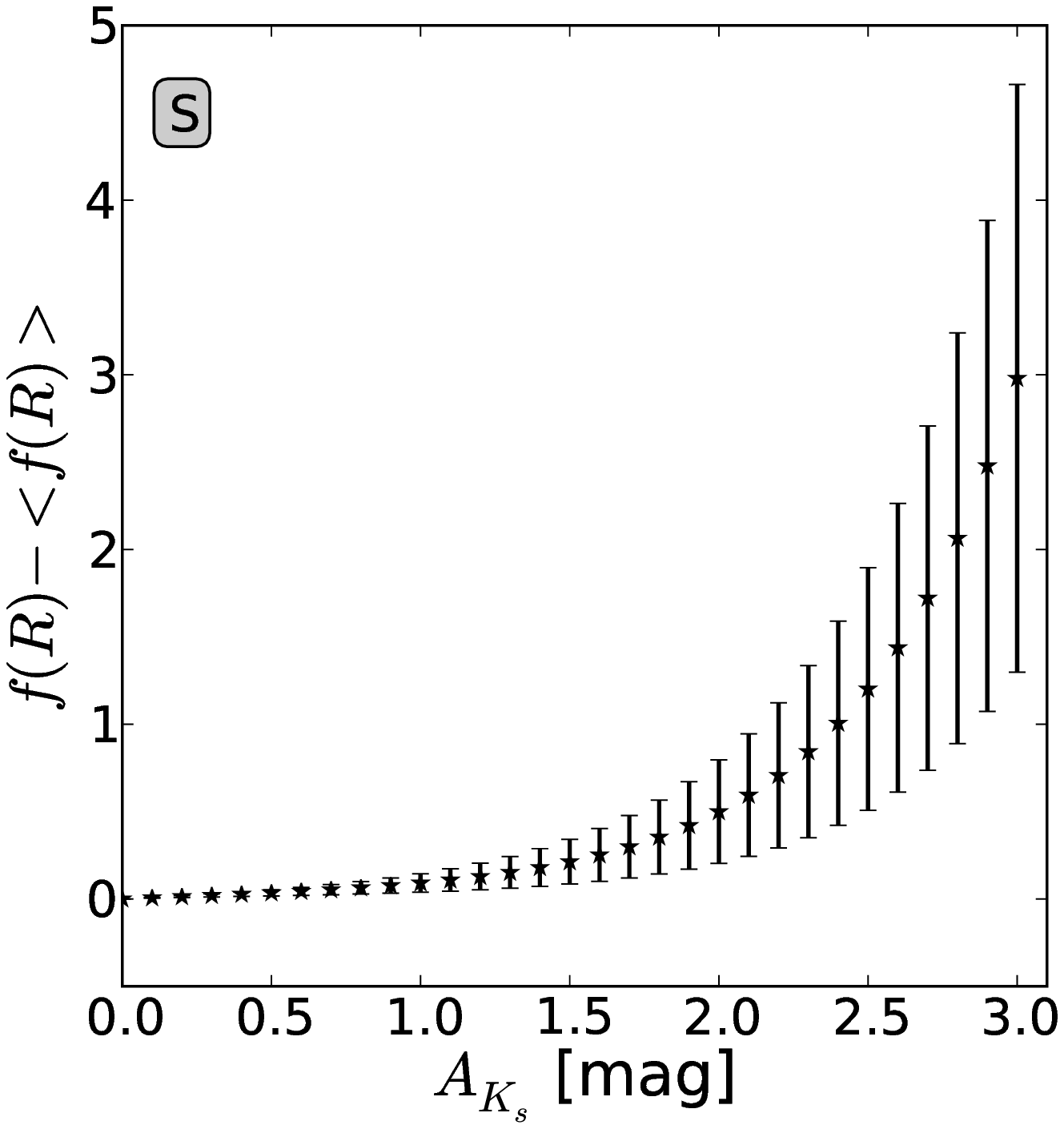}
\includegraphics[width=0.22\textwidth,height=0.22\textwidth]{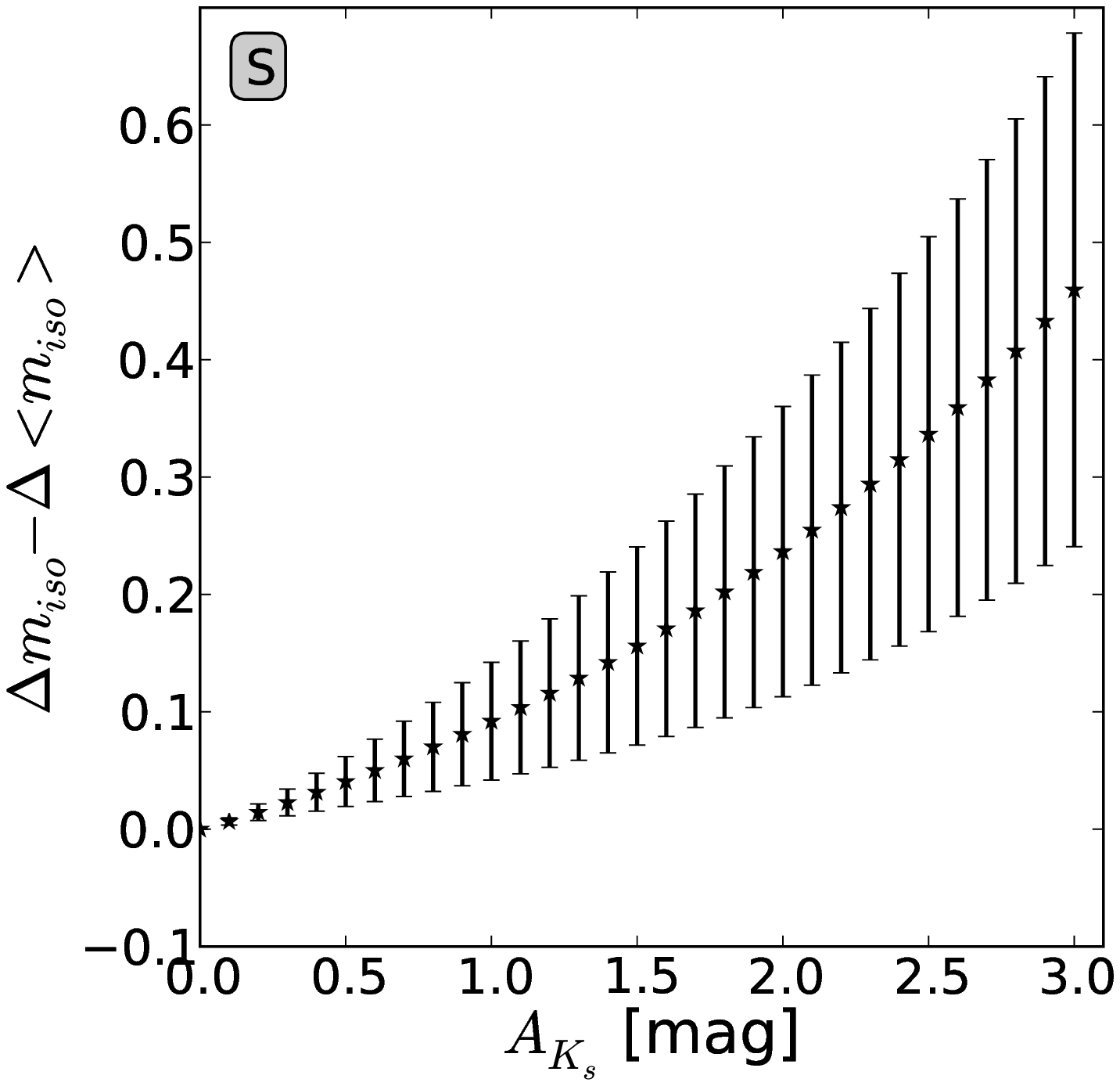}
\caption{Comparison between simulated corrections and average
corrections.}\label{avg-comp}
\end{center}
\end{figure}

To assess the accuracy of the average correction as a function of extinction, we made a
comparison between the simulated corrections and those derived from the average correction
in the extinction range $A_{K_s}=0\fm0-3\fm0$. The results are plotted in Fig.
\ref{avg-comp}. A summary is given in Table \ref{tavg}, where we list the mean difference
between the simulated data and the average correction value at the extinction levels
$A_{K_s}=0\fm5,\;1\fm0,\;2\fm0$ and $3\fm0$, as well as their scatter.
\setlength{\tabcolsep}{6pt}
\begin{table*}
\begin{minipage}{130mm}
\caption{Comparison between simulated corrections and the average corrections at
extinction levels $A_{K_s}=~0\fm5, 1\fm0$, $2\fm0$ and $3\fm0$.}\label{tavg}
\begin{center}
\begin{tabular}{llrrrrrrrr}
\hline
Galaxy&Param.&$0\fm5\;\;$&$SDE\;$&$1\fm0\;\;$&$SDE\;$&$2\fm0\;\;$&$SDE\;$&$3\fm0 \;\;$&$SD
E\;$\\
\hline
{\bf E/S0}&$f(R)$&0.0085&0.0081&0.0108&0.0170&0.0000&0.0450&$-0.0457$&0.1089\\
{\bf E/S0}&$\Delta
m_{iso}$&0.0425&0.0046&0.0045&0.0108&$-0.0020$&0.0266&$-0.0126$&0.0501\\
{\bf S}&$f(R)$&0.0348&0.0177&0.0913&0.0530&0.4994&0.2961&2.9797&1.6339\\
{\bf S}&$\Delta m_{iso}$&0.0406&0.0212&0.0920&0.0502&0.2365&0.1236&0.4594&0.2188\\
\hline
\end{tabular}
\end{center}
Average corrections were calculated using Eqns. \ref{fit1}, \ref{fit2}, and the parameters
in Table \ref{tavgfit}
\end{minipage}
\end{table*}

Similar to the optimized corrections, the average correction method performs better for
lower levels of extinction. Comparing Figs. \ref{tune-comp}, \ref{ctune-comp} and
\ref{avg-comp}, Tables \ref{ttune}, \ref{tctune} and \ref{tavg}, we notice that the
optimized corrections are more accurate than the average correction method.

\section{Discussion}
In this paper we present two methods to correct the isophotal magnitudes and radii for
galaxies observed in the NIR obscured by foreground extinction. The optimized correction
requires knowledge of the galaxy's light profile. The use of $\mu_c$ to estimate the
corrections is useful as it can also be independently used to roughly categorize the
galaxies as early or late type. The average correction method is more straightforward. It
gives average corrections at each extinction level in the $J$, $H$, $K_s$ observed
wavebands, and only requires the classification of a galaxy as early or late type.

To compare the accuracy of the average and the $\mu_0$ method, we used their comparison
with the simulated corrections, see Figs. \ref{tune-comp} and \ref{avg-comp} and Tables
\ref{ttune} and \ref{tavg}. For spiral galaxies estimating the magnitude corrections using
$\mu_0$ optimization, $\Delta m_{iso}(\mu_0)$ shows a systematic shift $\Delta=-0\fm01$
with a $SDE=0\fm0$ for an obscuration level of $A_{K_s}=0\fm5$. Meanwhile using the
average correction shows a $\Delta=0\fm04$ with $SDE=0\fm02$ at the same level of
extinction. At $A_{K_s}=1\fm0$, $\Delta m_{iso}(\mu_0)$ has a systematic shift of
$\Delta=-0\fm01$ and a $SDE=0\fm02$, while the average correction has a $\Delta=0\fm09$
with a $SDE=0\fm05$. Radius corrections for spiral galaxies revealed a similar trend for
the two methods, (see Tables \ref{ttune} and \ref{tavg}). The comparisons for the
elliptical galaxies are also given in Tables \ref{ttune} and \ref{tavg}.

Figures \ref{tune-comp}, \ref{avg-comp} and Tables \ref{ttune}, \ref{tavg} give the
comparison between the optimized and average corrections as compared to the simulated
corrections for the extinction values $A_{K_s}=~0\fm5,\;1\fm0,\;2\fm0,\;3\fm0$. They
clearly emphasize that the $\mu_0$ optimized correction is more accurate compared to the
average correction method. But the average correction remains more useful when applying
the corrections to galaxy parameters from a galaxy catalogue.

The $\mu_c$ optimized correction shows larger shifts and more dispersion than the $\mu_0$
optimized correction, but smaller shifts and tighter dispersion compared to the average
correction.

In the following we give some average correction values to correct magnitudes and radii of
obscured galaxies. The average correction estimates a $0\fm13$ correction to the isophotal
magnitude of elliptical galaxies at $A_{K_s}=1\fm0$. This magnitude correction is over and
above the $A_{K_s}=1\fm0$ correction. The magnitude correction shows a systematic shift of
$\Delta=0\fm01$ with a  $SDE=0\fm01$ at $A_{K_s}=1\fm0$. As a function of radius,
ellipticals are estimated to be $28.3\%$ smaller in radius at $A_{K_s}=1\fm0$, when using
the average correction. The radius correction shows a $\Delta=0.6\%$ and $SDE=0.9\%$ at
the same extinction level (see Table \ref{tavg1}).

The isophotal magnitudes of spiral galaxies at $A_{K_s}=1\fm0$ are expected to be $0\fm20$
brighter when applying the average correction. They show a systematic shift of
$\Delta=0\fm09$ with a $SDE=0\fm05$ at $A_{K_s}=1\fm0$. The average corrections predict
that spiral galaxies appear $28.4\%$ smaller at $A_{K_s}=1\fm0$. The corrections show a
systematic shift of $\Delta=4.4\%$ with a $SDE=2.6\%$ at the same obscuration level (see
Table \ref{tavg1}). The table also lists the expected magnitudes and radii corrections
respectively at the extinction levels $A_{K_s}=0\fm5,\;1\fm0$ and $2\fm0$. The table also
lists the systematic shifts and the $SDE$. The magnitude values listed in Table
\ref{tavg1} give the additional dimming, the galaxy size reduction, the systematic shift
when using the average correction and their respective $SDE$. The positive systematic
shifts indicate that the average correction under estimates the obscuration corrections.

It is worth mentioning that our corrections agree well with corrections given by
\citet{taka} for galaxies in the $K_s$ band obscured by $A_{K_s}\lesssim1\fm0$. In their
work they estimated the extra dimming to elliptical galaxies to be $\Delta m_{iso}=0\fm15$
compared to $\Delta m_{iso}=0\fm13$ as expected by our average correction. For spiral
galaxies they estimated the correction to be $\Delta m_{iso}= 0\fm18$ which also agree
well with our expected correction of $\Delta m_{iso}=0\fm20$. Compared to \citet{taka}
corrections, our results are useful for higher extinction levels in the three NIR bands
$J$, $H$ and $K_s$ i.e $A_{J,H,K_s}\leq3\fm0$.
\setlength{\tabcolsep}{4pt}
\begin{table*}
\begin{minipage}{140mm}
\begin{center}
\caption{Average magnitudes and radii corrections at
$A_{K_s}=0\fm5,\;1\fm0,\;2\fm0\;3\fm0$}\label{tavg1}
\begin{tabular}{lcccccccccccc}
\hline
&\multicolumn{3}{c}{$0\fm5$}&\multicolumn{3}{c}{$1\fm0$}&\multicolumn{3}{c}{$2\fm0$}
&\multicolumn{3}{c}{$3\fm0$}\\
\hline
[mag]&$\Delta m_{iso}$&$\Delta\;\;$&$SDE$&$\Delta m_{iso}$&$\Delta\;\;$&$SDE$&$\Delta
m_{iso}$&$\Delta\;\;$&$SDE$&$\Delta m_{iso}$&$\Delta$&$SDE$\\
\hline
{\bf E/S0}
&0.044&0.004&0.005&0.133&0.005&0.011&0.401&$-$0.002&0.027&0.766&$-$0.013&0.050\\
{\bf S}\hspace{4mm}
&0.061&0.041&0.021&0.198&0.092&0.050&0.648&$\;\;\;$0.236&0.124&1.296&$\;\;\;$0.459& 0.219
\\
\hline
\hline
[$\%$] &$f(R)$&$\Delta$&$SDE$&$f(R)$&$\Delta$&$SDE$&$f(R)$&$\Delta$&$SDE$&$f(R)$&$\Delta$&
$SDE$\\
\hline
{\bf E/S0}&13.0&0.7&0.6&28.3&0.6&0.9&54.8&0.0&0.9&73.3&$-$0.3&0.8\\
{\bf S}\hspace{4mm}&11.3&2.7&1.4&28.4&4.4&2.6&60.6&6.5&4.1&81.7&$\;\;\;$6.5&4.3\\
\hline
\end{tabular}
\end{center}
The table lists the extra dimming and the radii reduction as estimated by the average
corrections. It also lists the systematic shift $\Delta$ for comparing the simulated
correction and the average correction with the respective $SDE$. The positive systematic
shifts indicate that the average correction under estimate the extinction correction.
\end{minipage}
\end{table*}

\section{Conclusion}
We present two methods to correct galaxies for extinction in the $J$, $H$ and $K_s$ bands.
The optimized correction methods are more accurate than the average correction. However
the average correction method is more straightforward to apply as it requires no knowledge
of the light profile of the galaxy but only the classification of galaxies as early or
late types. The extinction corrections that we present here are considered as a NIR
extension to those for the $B$ band derived before by \citet{cameron}.

These corrections will be invaluable to the analysis of large scale structures in the
ongoing NIR galaxy survey along the Norma Wall in the ZOA. It will also be applicable to
other galaxy surveys e.g. 2MASX or prospective ESO galaxy surveys e.g. VISTA Kilo-Degree
Infrared Galaxy Survey (VIKING\footnote[1]{
http://www.eso.org/sci/observing/policies/PublicSurveys/sciencePublicSurveys.html}) using
the VISTA telescope.
\section*{Acknowledgments}
This publication makes use of data products from the Two Micron All Sky Survey, which is a
joint project of the University of Massachusetts and the Infrared Processing and Analysis
Centre/California Institute of Technology, funded by the National Aeronautics and Space
Administration and the National Science Foundation. The authors kindly acknowledge funding
from the South African National Research Foundation. IFR acknowledges the University of
Khartoum and the Stichting Steunfonds Soedanese Studenten for financial support.

The authors would also like to thank the referee for the very careful reading of the
manscript and the many remarks and comments they made to improve the overall work.

\end{document}